\documentclass[%
 aip,
 amsmath,amssymb,
 reprint,%
]{revtex4-2}

\usepackage{graphicx}
\usepackage{dcolumn}
\usepackage{bm}
\usepackage{caption}
\usepackage{subcaption}
\usepackage[utf8]{inputenc}
\usepackage[T1]{fontenc}
\usepackage{mathptmx}
\usepackage{adjustbox}
\usepackage{hyperref}
\usepackage{pifont}

\usepackage{listings}
\usepackage[dvipsnames]{xcolor}

\definecolor{codegreen}{rgb}{0,0.6,0}
\definecolor{codegray}{rgb}{0.5,0.5,0.5}
\definecolor{codepurple}{rgb}{0.58,0,0.82}
\definecolor{backcolour}{rgb}{0.95,0.95,0.92}
\definecolor{color1}{rgb}{0.08523511613408935,0.32661779003565533, 0.2973201282529313}
\definecolor{color2}{rgb}{0.6328422475018423,0.4747981096220677, 0.29070209208025455}
\definecolor{color3}{rgb}{0.7779565181455343,0.7069421942599752, 0.9314406084043191}

\definecolor{color11}{rgb}{0.08605633600581405,0.23824692404212, 0.30561236308077167}
\definecolor{color12}{rgb}{0.32927729263408284,0.4762845556584382, 0.1837155549758328}
\definecolor{color13}{rgb}{0.6328422475018423,0.4747981096220677, 0.29070209208025455}
\definecolor{color14}{rgb}{0.8299576787894204,0.5632024035248271, 0.7762744444444445}

\lstdefinestyle{mystyle}{
    backgroundcolor=\color{backcolour},   
    commentstyle=\color{codegreen},
    keywordstyle=\color{magenta},
    numberstyle=\tiny\color{codegray},
    stringstyle=\color{codepurple},
    basicstyle=\ttfamily\footnotesize,
    breakatwhitespace=false,         
    breaklines=true,                 
    captionpos=b,                    
    keepspaces=true,                 
    numbers=left,                    
    numbersep=5pt,                  
    showspaces=false,                
    showstringspaces=false,
    showtabs=false,                  
    tabsize=2
}

\usepackage{amsmath}
\usepackage{comment}

\usepackage{hyperref}
\usepackage{cleveref}

\newcommand\myshade{85}
\colorlet{mylinkcolor}{violet}
\colorlet{mycitecolor}{YellowOrange}
\colorlet{myurlcolor}{Aquamarine}

\newcommand{\ud}{\textrm d}
\newcommand{\ue}{\textrm e}
\newcommand{\ui}{\textrm i}

\hypersetup{
  linkcolor  = mylinkcolor!\myshade!black,
  citecolor  = mycitecolor!\myshade!black,
  urlcolor   = myurlcolor!\myshade!black,
  colorlinks = true,
}

\hypersetup{
  linkcolor  = mylinkcolor!\myshade!black,
  citecolor  = mycitecolor!\myshade!black,
  urlcolor   = myurlcolor!\myshade!black,
  colorlinks = true,
}

\begin{document}

\preprint{AIP/123-QED}
\title[Neural Networks for wall pressure spectral models]{Artificial Neural Networks Modelling of Wall Pressure Spectra Beneath Turbulent Boundary Layers}

\author{J. Dominique}
 \email{joachim.dominique@vki.ac.be}
\author{J. Van den Berghe}%
 \author{C. Schram}%
\author{M. A. Mendez}%
\affiliation{ 
von Karman Institute for fluid dynamics. Chaussée de Waterloo 72, 1640 Rhode-Saint-Genèse (Belgium)
}

\date{\today}

\begin{abstract}

We analyse and compare various empirical models of wall pressure spectra beneath turbulent boundary layers and propose an alternative machine learning approach using Artificial Neural Networks (ANN). The analysis and the training of the ANN are performed on data from experiments and high-fidelity simulations by various authors, covering a wide range of flow conditions. We present a methodology to extract all the turbulent boundary layer parameters required by these models, also considering flows experiencing strong adverse pressure gradients. Moreover, the database is explored to unveil important dependencies within the boundary layer parameters and to propose a possible set of features from which the ANN should predict the wall pressure spectra. The results show that the ANN outperforms traditional models in adverse pressure gradients, and its predictive capabilities generalise better over the range of investigated conditions. The analysis is completed with a deep ensemble approach for quantifying the uncertainties in the model prediction and integrated gradient analysis of the model sensitivity to its inputs. Uncertainties and sensitivities allow for identifying the regions where new training data would be most beneficial to the model's accuracy, thus opening the path towards a self-calibrating modelling approach. The ANN model has been made publicly available at \url{https://github.com/DominiqueVKI/VKI_researchWPS}.

\end{abstract}

\maketitle

\section{Introduction}

Wall pressure fluctuations underneath a turbulent boundary layer (TBL) induce fluid-structural coupling, fatigue-induced component failure \cite{hambric2004vibrations} and vibro-acoustic noise transmission in many applications including air and ground transport, or wind turbine blades \cite{bull1996,Avallone2018,Tang2019,roger2005back}. In each case, the efficiency of the aero-vibro-acoustic coupling mechanism depends on the amplitude of the pressure fluctuations, on their spatial coherence, and on the velocity at which the pressure footprint of the TBL vortical structures is advected. Reliable engineering models are still needed for those characteristics, in particular when the TBL is subjected to a streamwise pressure gradient.

Considerable research effort has been devoted to the measurement and simulation of the above quantities for relatively canonical flows such as flat plate boundary layers in null \cite{blake1970} or moderate adverse pressure gradients \cite{keith1991}, or over airfoils \cite{moreau2005}. High fidelity simulations such as LES\cite{cohen2018} or DNS\cite{choi1990} can provide an unprecedented amount of detailed data about the unsteady flow field but remain too costly for the design stage in many engineering applications. Wind tunnel experiments\cite{farabee1991,van2018} share the same problem.

This explains the sustained interest in simplified, empirical or semi-empirical models, relating the wall pressure fluctuations, their spectrum and coherence lengths, to the main characteristics of a TBL\cite{bull1996}. Following the work of Kraichnan \cite{Kraichnan1956} and Panton \& Linebarger \cite{panton1974}, Blake \cite{Blake_1986} derived models based on the integration of the Poisson equation in the wavenumber-frequency domain. This method requires the vertical distribution of the mean longitudinal velocity and the wall-normal Reynolds stresses, and leads to high dimensional integrals, which can be difficult to compute\cite{grasso2019}.

A different approach was followed by Amiet \cite{Amiet1976} and Chase \& Howe \cite{Howe_1998} for the prediction of airfoil trailing-edge noise. It makes use of empirical or semi-empirical relations to relate the wall pressure spectra to statistical parameters of the boundary layer such as its thickness, external velocity or friction velocity. This procedure is attractive because the TBL parameters can be computed from low-CPU Reynolds Averaged Navier Stokes (RANS) simulations combined with more or less complex correlations.

Several improvements have thus been brought along this line over the last couple of decades. The empirical model of Chase-Howe was modified by Goody \cite{goody2004} to account for Reynolds number effects and improve the high-frequency behaviour. This model accurately predicts wall pressure spectra in the absence of pressure gradients, as demonstrated by Hwang \cite{Hwang2007}. Extensions of this last approach to account for the presence of pressure gradients were proposed by Kamruzzaman \cite{kamruzzaman2015}, Rozenberg \cite{rozenberg2012}, Hu \cite{Hu2018}, or Lee \cite{lee2018}.

Nevertheless, these models have proved valid only within specific ranges of the pressure gradient. Although rooted in the physics of the boundary layer, their derivation remains somewhat ad-hoc and usually requires the data-driven calibration of several coefficients\cite{catlett2016}. This process is essentially a regression problem, in which the definition of the parametric function is driven by physical considerations, scaling laws and modeller's ingenuity. Furthermore, for boundary layers in non-equilibrium the local turbulence dynamics and associated wall pressure fluctuations cannot be characterized exclusively using local parameters such as Clauser's parameter $\beta$ and the friction Reynolds number $Re_\tau$, but requires additional parameters to account for the boundary layer history\cite{Bobke2017} .

Given those difficulties, it becomes tempting to alleviate the problem of elucidating the complex dependencies between the TBL physical parameters, and to consider machine learning techniques instead, with their fast-growing arsenal of tools to solve regression problems\cite{Brunton2020}. Methods and tools from genetic programming\cite{Banzhaf1997} and deep learning\cite{Goodfellow} allow for deriving complex nonlinear functions from data, ultimately shifting the modelling challenge from the definition of the parametric function to the definition of an appropriate set of input parameters.

In this line, the present authors have recently proposed Genetic Programming to model pressure wall spectra \cite{dominique2021}. Genetic Programming is a collection of evolutionary techniques that automatically derive and calibrate analytic expressions linking input and output variables. While providing encouraging results (also reported in Section~\ref{sec:3E} of this article), the procedure requires a significant amount of time to converge when the parameter space formed by the number of operators and input functions becomes large\cite{cranmer2020}. On the other hand, limiting the set of candidate gene structures requires a large amount of knowledge about the regression task at hand or a tedious trial-and-error procedure.

Deep learning is considered in this work as an alternative. Deep learning is a subset of machine learning concerned with training large Artificial Neural Networks (ANN). These are recursive and distributed architectures of simple computational units, called neurons. ANNs are known as \emph{universal function approximators} because of their ability to approximate \emph{any} function\cite{Hornik1989}. This paper presents a wall pressure spectrum model based on ANNs. The model was calibrated (or \emph{trained}, in the machine learning terminology) on a wide range of conditions using experimental and numerical data from various authors. The model's predictive capabilities are compared with classic semi-empirical models, and model uncertainties are derived using an ensemble approach\cite{Abdar2021,Gawlikowski2021}. The ANN model and the uncertainty predictor have been made publicly available at \url{https://github.com/DominiqueVKI/VKI_researchWPS}.

The remaining of this paper is structured as follows. Section \ref{sec:2} presents the modelling problem, introducing the boundary layer parameters of existing wall pressure spectra models, described in Section \ref{sec:3}.  Section \ref{sec:4} presents the numerical and experimental test cases that were used to calibrate the model in a first stage, and to validate it afterwards. Section \ref{sec:5} presents the methodology used to extract the boundary layer parameters, to design the ANN architecture to compute its uncertainty. Results are presented and discussed in Section \ref{sec:6} where we compare the prediction capabilities of the models on the present dataset. Finally, Section \ref{sec:7} collects conclusions and perspectives.

\section{Boundary layer wall pressure spectrum: definitions and parameters}
\label{sec:2}

\begin{figure}[h]
\includegraphics[width=0.45\textwidth]{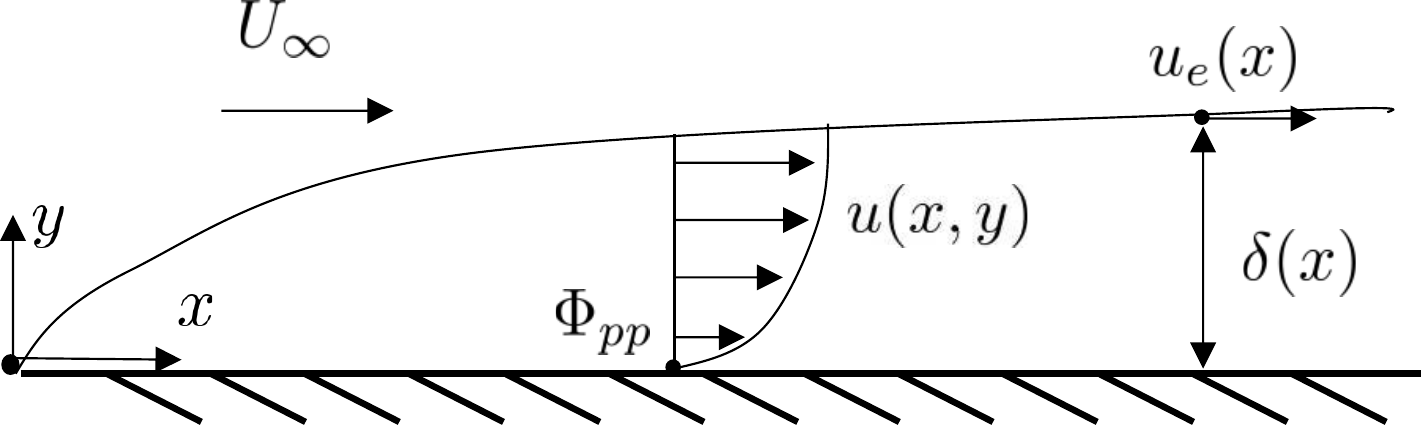}
\caption{Boundary layer developing over a large flat plate}
\label{fig:2dbl}
\end{figure}

The configuration of interest is sketched in Fig. \ref{fig:2dbl}. A two-dimensional boundary layer of thickness $\delta(x)$ develops as a flow with uniform velocity $U_\infty$ runs over a plate (or an airfoil). The fluid density and kinematic viscosity are denoted as $\rho$ and $\nu$, respectively. Because our focus is on turbulent flows, all velocities are to be interpreted as \emph{mean} velocities. Moreover, we consider conditions with negligible compressibility effects (i.e. with $Ma<0.3$, with $Ma=U_\infty/c_0$ the Mach number and $c_0$ the speed of sound).

The velocity of the flow at the edge of the boundary layer is the equilibrium velocity $u_e$. The boundary layer thickness is customarily defined as the distance from the wall at which the stream-wise velocity is $u(x,\delta)=0.99 U_\infty=u_e(x)$. This is the definition considered in this work for all the experimental datasets. An alternative definition is based on the pseudo-velocity\cite{deuse2019} $u^*(x,\delta)=0.99 u^*(x,\infty) = u_e(x)$, defined as

\begin{equation}
        \label{pseudo}
        u^*(x,y)=-\int^{y}_{0}\Omega_z(x,\xi) d\xi\,\,
\end{equation} where $\xi$ is a dummy integration variable along $y$ and $\Omega_z$ is the third component of the (mean) vorticity $\Omega=\nabla \times \mathbf{u}$, with $\mathbf{u}=(u,v)$ the (mean) velocity vector.

This quantity becomes constant when vorticity vanishes outside the boundary layer. The definition of $\delta$ based on the psuedo-velocity was used in this work for all the numerical datasets, as it was found to be more accurate in regions where the outer flow does not have uniform velocity.

Within the boundary layer region ($y<\delta$), the vertical pressure gradient is $\partial_y p  =0$, while the acceleration of the flow in the outer region ($y>\delta$) can produce an Adverse Pressure Gradient (APG, $\partial_x p>0$), or a Favorable Pressure Gradient (FPG, $\partial_x p<0$). In the absence of flow acceleration, i.e. $\partial_x u_e=0$, one has a Zero Pressure Gradient (ZPG) condition.

The point-wise power spectral density of the pressure fluctuations can be obtained from the Fourier transform of the auto-correlation function:
\begin{equation} 
\label{eq:Phipp}
    \Phi_{pp}(x,\omega) = \int_{-\infty}^{+\infty} R(x,\tau) \, \ue^{\scriptsize -\ui\omega \tau} \, \ud\tau,
\end{equation}
where $\ui=\sqrt{-1}$, $\omega=2 \pi f$ is the angular pulsation and 
\begin{equation}
    \label{R}
R(x,\tau) = \int_0^\infty p'(x,t) \, p'(x,t+\tau) \, \ud t
\end{equation}
is the auto-correlation of the pressure fluctuations $p'$ at the wall ($y=0$), here assumed to be invariant along the spanwise direction $z$. 

Empirical models aim at linking the space dependency of $\Phi_{pp}(x,\omega)$ to local boundary layer quantities. These include integral parameters measuring the boundary layer thickness and the expected averaged velocity profile, as well as others accounting for the influence of the pressure gradient. 
Besides $\delta$, the boundary layer thickness can be measured also in terms of the displacement thickness $\delta^*$ and momentum thickness $\theta$:
\begin{subequations}
\label{eq:thickness}
\begin{gather}
\delta^*(x) = \int_0^{\delta}\left( 1 - \frac{u(x,y)}{u_e} \right) dy\,,\\
  \theta(x) = \int_0^{\delta} \frac{u(x,y)}{u_e}\left( 1 - \frac{u(x,y)}{u_e} \right) dy\,.
\end{gather}
\end{subequations} 
These account, respectively, for the mean flux and the momentum flux deficits due to the boundary layer's velocity profile $u(y)$. In a ZPG, the velocity profile is excepted to obey a universal form\cite{coles1956}: 

\begin{equation}
    \label{eq:universal_vel}
    u^+(y^+) =
    \begin{cases}
      \mathcal{F}_w (y^+) + \frac{\Pi}{\kappa} \mathcal{W}\left(\eta\right) & \text{if } y < \delta \\
      u_e/u_\tau & \text{if } y \geq \delta
     
     \end{cases}
\end{equation} where $u^+=u/u_\tau$, with $u_\tau=\sqrt{\tau_w/\rho}$ the friction velocity and $\tau_w$ the wall shear stress, $y^+=y/\delta_v$, with $\delta_v=\nu/u_\tau$ the viscous length scale and $\eta=y/\delta$ a dimensionless coordinate which is better suited than $y^+$ far from the wall. The first term is the \emph{law of the wall}; the second term is a corrective term introduced by Coles\cite{coles1956} and is known as \emph{wake function}. The coefficient $\Pi$ is the wake strength parameter and $\kappa$ is the von Karman constant. We return on the definition of the velocity profiles wall function $\mathcal{F}_w (y^+)$ and the wake function $\mathcal{W}$ in Section \ref{sec:5p1}. For the moment, it suffices noticing that, as shown by Nagib and Chauhan \cite{nagib}, the impact of a pressure gradient can be accounted for while still keeping the ansatz in \eqref{eq:universal_vel} and modifying the functions $\mathcal{F}_w$ and $\kappa$ according to the strength and the sign of the pressure gradient.

From the set of parameters introduced thus far, empirical models for the wall pressure spectra search for a relation of the form:
\begin{equation}
\label{Model_Phi}
    \Phi_{pp}(\omega)=f\left(\delta,\delta^*,\theta, u_e, \rho, \nu, \tau_w, \partial_x p, c_0, \Pi \right) \,,
\end{equation}
having omitted the functional dependency on the space coordinate $x$, which is included in the boundary layer parameters. It is worth noticing that the parameters $\delta^*$ and $\theta$ could be deduced from the model of the velocity profile in \eqref{eq:universal_vel} and thus one could postulate a relation of the form $\delta^*,\theta=g(\delta,u_e,\rho,\nu, \tau_w,\partial_x p, c_0)$. Nevertheless, in this work we use a model for $u(y)$ (as in equation \ref{eq:universal_vel}) only for the computation of $\tau_w$ and $\Pi$ and consider $\delta^*$ and $\theta$ as independent inputs to the correlation to be derived.

Because only three fundamental units are involved, the Buckingham-Pi theorem suggests that such a relation can be conveniently written in dimensionless form as 
\begin{equation}
    \frac{\Phi_{pp} u_e}{\delta^* \tau_w^2}\Biggl(\frac{\omega \delta^*}{u_e}\Biggr)=f\left(\frac{\delta}{\delta^*}, \frac{\theta}{\delta^*}, \frac{c_0}{u_e}, \frac{\rho u_e^2}{\tau_w}, \frac{\nu \delta^*}{u_e}, \frac{\delta^*}{\tau_w}\frac{dp}{dx}, \Pi\right) \,,
\end{equation} having used $\delta^*$, $\tau_w$ and $u_e$ as repeated variables in the scaling. Adjusting the derived dimensionless number to those more commonly encountered in the literature of wall pressure spectra, the dimensionless function to be derived is thus
\begin{equation} \label{eq:Buckingham}
\tilde{\Phi}_{pp}(\tilde{\omega}) = \tilde{f}\left(\Delta, H, Ma, \Pi, C_f, R_T, \beta \right),
\end{equation} where $\tilde{\Phi}_{pp}=\Phi_{pp}/(\delta^* \tau_w^2/u_e)$ is the dimensionless power spectral density, $\tilde{\omega}=\omega/(u_e/\delta^*)$ is the dimensionless frequency, $\Delta=\delta/\delta^*$ is the Zagarola-Smits's parameter \cite{zagarola}, $H=\delta^*/\theta$ is the boundary layer's shape factor, $Ma=u_e/c_0$ is the Mach number, $C_f=\tau_w/(\rho u_e^2)$ is the friction coefficient, $R_T=(\delta^*/u_e)/(\nu/u_\tau^2)$ is the time scale ratio, also known as the Reynolds number effect of the pressure spectrum \cite{goody2004}, and $\beta = (\theta/\rho u_\tau^2) \partial_x p $ is the Clauser parameter. 

Finally, it is worth noticing that the scaling laws in \eqref{eq:Buckingham}, and particularly the definition of the dimensionless frequency and the wall pressure spectra, are known to be reasonable for the low and the mid-frequency range ($\omega \nu / u_\tau^2 < 100$) but not for the higher frequency range \cite{blake1986}, in which a scaling based on the inner variables, i.e. $\hat{\omega}=\omega \nu/(u_\tau^2)$, is more appropriate\cite{bull1996}. Nevertheless, the lack of universal scaling for all frequencies is compensated by the Reynolds number $R_T$, which is the ratio of the two leading time scales in the spectrum.

\section{Empirical models}
\label{sec:3}

We now introduce various forms of Eq.~\eqref{eq:Buckingham} from five models available in the literature.  These are used to benchmark the ANN model derived in this work.

\subsection{Goody's ZPG model based on self-similarity}

The empirical wall pressure spectra proposed by Goody\cite{goody2004} was built based on the principle of scaling and self-similarity. The primary precept is that the wall pressure spectra are proportional to different scales at low and high frequencies. The low-frequency scales with outer flow variables ($\delta^*$ and $u_e$) and the high-frequency scales with the viscosity and inner variables ($\nu$ and $u_\tau$). Such scaling is similar to the two-layer scaling of turbulent boundary layer velocity profile using a viscous sublayer and a logarithmic layer \cite{coles1956}. 

Based on this principle, Goody further extended the model of Chase and Howe \cite{Howe_1998} to account for this outer-to-inner scaling ratio by including the Reynolds number, with the following expression:
\begin{equation*}
\label{eq:Goody}
\frac{\Phi_{pp}(\omega) u_e}{\tau_w^2 \delta} = \frac{ C_2(\omega \delta / u_e)^2 }{\left( (\omega \delta / u_e)^{0.75} + C_1\right)^{3.7} + \left( C_3(\omega \delta / u_e) \right)^7},
\end{equation*} where $C_1 = 0.5$, $C_2=3.0$, and $C_3 = 1.1 R_{T,\delta}^{-0.57}$ describes the effect of the Reynolds number, which these author define with respect to $\delta$ and not $\delta^*$, i.e. $R_{T,\delta}=(\delta/u_e)/(\nu/u_\tau^2)$. This model, tuned using the constants listed above, describes accurately wall pressure spectra beneath a turbulent boundary layer in the absence of pressure gradients, as demonstrated by Hwang \cite{hwang2009comparison}.

\subsection{Rozenberg's extension for APG}

Rozenberg \emph{et al.}\cite{rozenberg2012} extended Goody's model by including the effects of adverse pressure gradients. The authors assumed these could be incorporated into the model using the Clauser parameter $\beta$ and the boundary layer wake parameter $\Pi$. This assumption originated from an analogy with the scaling of turbulent boundary layer velocity profiles, which requires a wake correction.  
Rozenberg's model has three main extensions of Goody's model. First, the slope of the mid-frequency range is allowed to be a function of the pressure gradient. Second, the global level of pressure fluctuation is allowed to increase with $\beta$ and $\Pi$. Third, the reference length used for the scaling of the model is changed to the displacement thickness.

The resulting model is:
\begin{equation*}
\label{eq:Rozenberg}
\frac{\Phi_{pp}(\omega) u_e}{\tau_w^2 \delta^*} = \frac{\Bigg[ 2.82\Delta^2(6.13\Delta^{-0.75}+F_1)^{A_1} \Bigg] \left[4.2 \frac{\Pi}{\Delta} + 1\right] \tilde{\omega}^2}{[4.76 \tilde{\omega}^{0.75} + F_1]^{A_1} + (C_3\tilde{\omega})^{7}},
\end{equation*}  where $F_1 = 4.76\left({1.4}/{\Delta}\right)^{0.75}[0.375A_1-1]$, $A_1 = 3.7 + 1.5\beta$ and $C_3 = 8.8R_{T,\delta}^{-0.57}$. This formulation falls back on Goody's for ZPG. However, this model was only calibrated on APG flows data and is therefore unable to predict FPG. 

\subsection{Kamruzzaman's extension for FPG}

Kamruzzaman \emph{et al.}\cite{kamruzzaman2015} extended Rozenberg's model to a larger database, including APG and FPG conditions. 
Differently from the previous works, the authors reduce the set of input parameters by linking Clauser parameter to the wake coefficient via the relation $\Pi = 0.8 \left( \beta + 0.5 \right)^{0.75}$.  

The amplitude of the model predictions was slightly increased for both APG and FPG. The dependency of the Reynolds number suggested by Goody was also changed to best fit their database. As a result, the prediction of this model does not agree with Goody's model for ZPG. In opposition to Rozenberg's model, the variations of the slope in the mid-frequency range were not observed in Kamruzzaman's study. The resulting model is:

\begin{equation*}
\label{eq:Kamruzzaman}
\frac{\Phi_{pp}(\omega) u_e}{\tau_w^2 \delta^*} = \frac{B_2 \tilde{\omega}^2}{[\tilde{\omega}^{1.64} + 0.27]^{2.47} + [B_3 \tilde{\omega} ]^7},
\end{equation*} where $ B_2  = 0.45(1.75(\Pi^2\beta^2)^{0.5(H/1.31)^{0.3}}+15) $ and $ B_3 = (1.15R_T)^{-2/7}$. 

Due to the empirical relationship between the wake coefficient and the Clauser parameters, the model dependency on $\Pi$ is enslaved to its dependency on $\beta$. The shape factor $H$ is also used for the first time as a modelling tool to account for history effects of the boundary layer developments. The use of the shape factor to model history effects in wall pressure spectra was later used in 2018 by Hu \emph{et al.} \cite{Hu2018} who proposed another empirical wall pressure spectral model with no direct influence of the pressure gradient.

\subsection{Lee's combined model}

Lee \emph{et al.}\cite{lee2018} performed an extensive comparison between all the models previously discussed, and highlighted that although each of these performed well on their training dataset, the match with other data proved to be limited. Lee \emph{et al} proposed a combination of the previous models with the following modifications: \textit{i}) the transition from mid to high frequency ranges is made dependent on the pressure gradient, \textit{ii}) a correction of the spectrum amplitude at low frequencies for ZPG and \textit{iii}) a larger increase of the amplitude for large APGs.
\begin{equation*}
\label{eq:Lee}
\frac{\Phi_{pp}(\omega) u_e}{\tau_w^2 \delta^*} = \frac{B_2 \tilde{\omega}^2}{[4.76\tilde{\omega}^{0.75}+d^*]^e + [C_3\tilde{\omega}]^{h^*}},
\end{equation*} where $B_2 = max(a,(0.25\beta - 0.52 )a)$, $a=2.82\Delta^2(6.13\Delta^{-0.75}+d)^{3.7+1.5\beta}$, $d^*=max(1.0,1.5d)$, $d=4.76(1.4/\Delta)^{0.75}[0375e-1]$, $e=3.7 + 1.5\beta$, $C_3 = 8.8 R_{T,\delta}^{-0.57}$ and finally $h^* = min(3,0.139+3.1043\beta)+7$. 

Lee's model is close to the Goody model for ZPG flows with slight discrepancies at low frequencies. Despite its complexity and being non-continuous because of the minimum/maximum operators, this is nowadays known as one of the most accurate empirical models for FPG and APG boundary layers. However, as it will be shown in \textsc{Sec.} \ref{sec:6}, this model still faces difficulties in the presence of strong APG.

\subsection{Dominique's model obtained from Gene Expression Programming}
\label{sec:3E}

Dominique \emph{et al.}\cite{dominique2021} presented a data-driven methodology for empirical models of wall pressure spectra model. Unlike the models discussed thus far, this is not derived as an extension of previous models but is derived \textit{ex novo} using Gene Expression Programming \cite{ferreira2006} (GEP). 

The method was originally implemented using a single experimental dataset of flat plate boundary layer \cite{salze2014}, but it is extended in this work to the larger dataset described in Sec \ref{sec:4}. This approach returns the following empirical model:

\begin{equation*}
\label{eq:Dominique}
\frac{\Phi_{pp}(\omega) u_e}{\tau_w^2 \delta^*} = \frac{\left( 5.41+ C_f(\beta_C + 1 )^{5.41}\right) \tilde{\omega}}{\tilde{\omega}^2 + \tilde{\omega} + (\beta+1)Ma + (\tilde{\omega}+3.6)\frac{\tilde{\omega}^{4.76}}{C_f R_T^{5.83}}},
\end{equation*}

The GEP approach returns a model which is structurally close to the other models described in this section in that it assumes a rational function in $\tilde{\omega}$. However, some differences are evident at low frequencies and high frequencies. At low frequencies, the GEP model yields an asymptotic dependence as $\tilde{\omega}^{1}$, instead of the $\tilde{\omega}^{2}$ predicted by the Kraichnan-Phillips theorem~\cite{Kraichnan1956}. 
At high frequencies, the GEP model yields a decay as $\tilde{\omega}^{-4.76}$ instead of the $\tilde{\omega}^{-5}$ theoretically predicted by Blake\cite{blake1986}.

While these discrepancies might appear to violate physical models, we recall that these theoretical results are not always confirmed experimentally. For example, the theoretical slope at low frequencies has been rarely observed in wind tunnel measurements. One of the very few experimental confirmations, brought by Farabee and Casarella~\cite{farabee1991}, required low-speed measurements performed in a particularly quiet wind tunnel and dedicated noise cancellation techniques~\cite{goody2004}. 

The GEP somehow accounts for these effects and appears to have better predictive performances on realistic scenarios. The differences at high frequencies ($\tilde{\omega}^{-4.76}$ instead of $\tilde{\omega}^{-5}$) is somewhat minor if one considers that the GEP model is the result of a pure regression, unaware of physics-based insights.

\section{Experimental and numerical datasets used for training and validation}\label{sec:4}

The dataset analyzed in this work include the experimental database by Salze \emph{et al}\cite{salze2014} on the boundary layer over a flat plate and three high-fidelity numerical simulations of the flow over a controlled diffusion (CD) airfoil, by Deuse \emph{et al}\cite{deuse2019}, Hao \emph{et al}\cite{wu2018} and Christophe \emph{et al}\cite{christophe2009}.

The flat plate experimental dataset from Salze \emph{et al}\cite{salze2014} includes ZPG and moderate APG and FPG conditions. The first should be suited for Goody's model whereas the others should be within the range of validity of Lee's or Kamruzzaman's models. 

The CD airfoil is a benchmark problem for aero-acoustic noise prediction. This airfoil is representative of engine cooling axial fans for automotive or HVAC systems. It has a 4\% relative thickness and a leading-edge camber angle of $12^{\circ}$. In nominal operating conditions, the airfoil has a laminar boundary layer on the pressure side and a short separation bubble at the leading edge, on the suction side, followed by a reattachment and transition to a turbulent boundary layer.

In contrast with the flat plate case, the CD configuration provides a variety of conditions along the airfoil chord, as the development of the TBL continuously adapts to the local Reynolds number and pressure gradient. The TBL data have thus been extracted at a number of streamwise locations starting from the trailing edge (the relevant location for the prediction of trailing edge noise) and further upstream of it. A brief descrition of each dataset, as well as the precise locations at which each profiles are extracted are reported in the following subsections.

\subsection{Salze's flat plate TBL subjected to pressure gradients}

Salze\cite{salze2014} performed an experimental investigation of the wall pressure spectrum under moderately favourable and adverse pressure gradients in the closed test section wind tunnel of the University of Lyon. They used a rotating antenna of pinhole microphones to measure the wall pressure fluctuations and a sloped opposite wall to induce pressure gradients. All wall pressure spectra are obtained at the location of the antenna in the wind tunnel, changing the flow velocity.

\begin{figure}[h]
\includegraphics[width=0.5\textwidth]{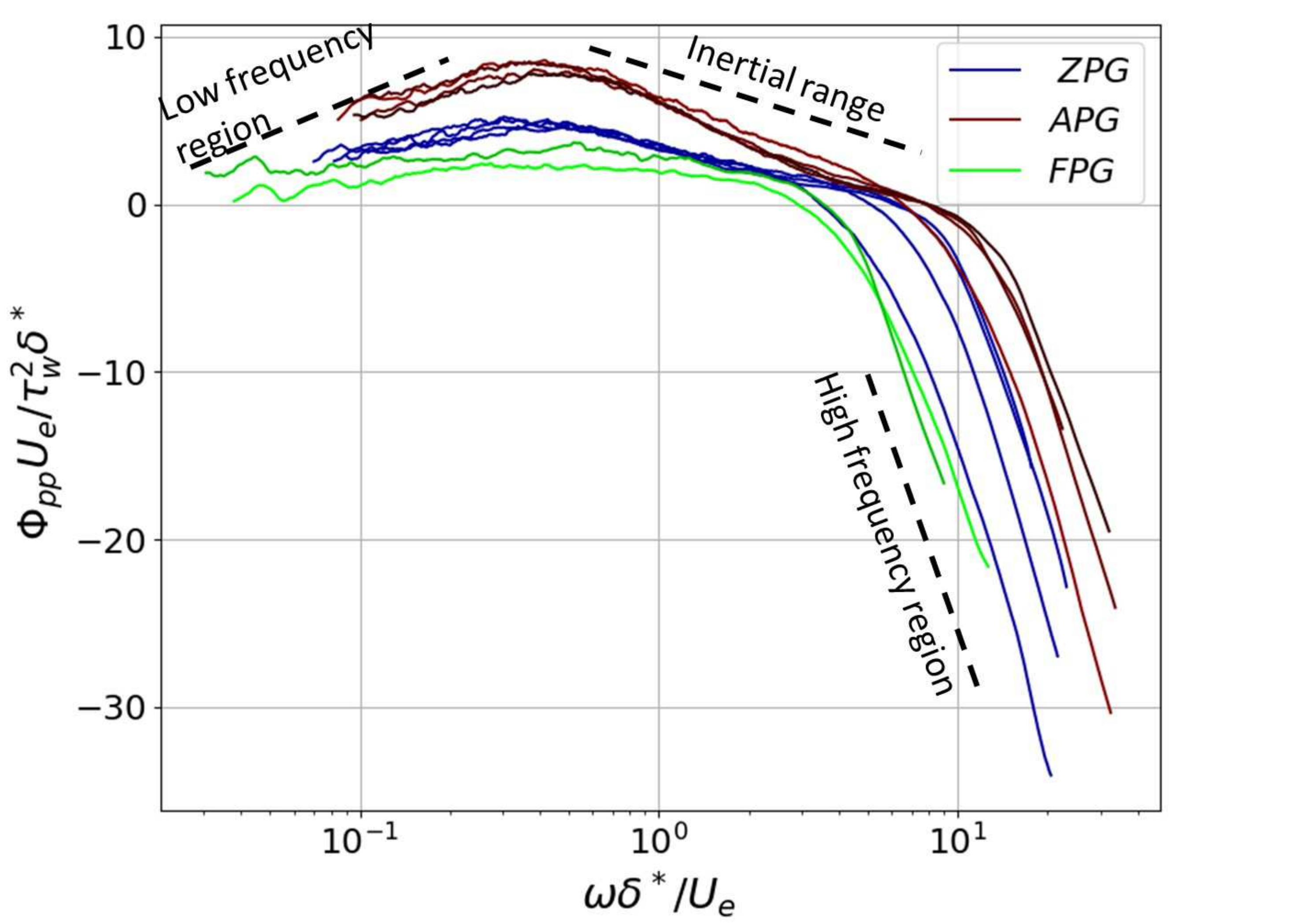}
\caption{Experimental wall pressure spectra from the flat plate TBL database of Salze\cite{salze2014}.}
\label{fig:salze_data}
\end{figure}

The measurements of ZPG boundary layers were found close to the expected Goody model. However, only moderate effects of the pressure gradient were observed. At low frequency, the amplitude of the spectra was slightly decreased for FPG and increased for APG. As it can be observed in Fig. \ref{fig:salze_data}, the pressure gradients also affect the inertial slope in the mid-frequency region. At high frequency, the spectrum decays as $\omega^{-5}$ for all pressure gradients.

\subsection{Deuse's CD airfoil TBL at 8 degrees of angle of incidence}

\begin{figure*}
\centering
\includegraphics[width=\textwidth]{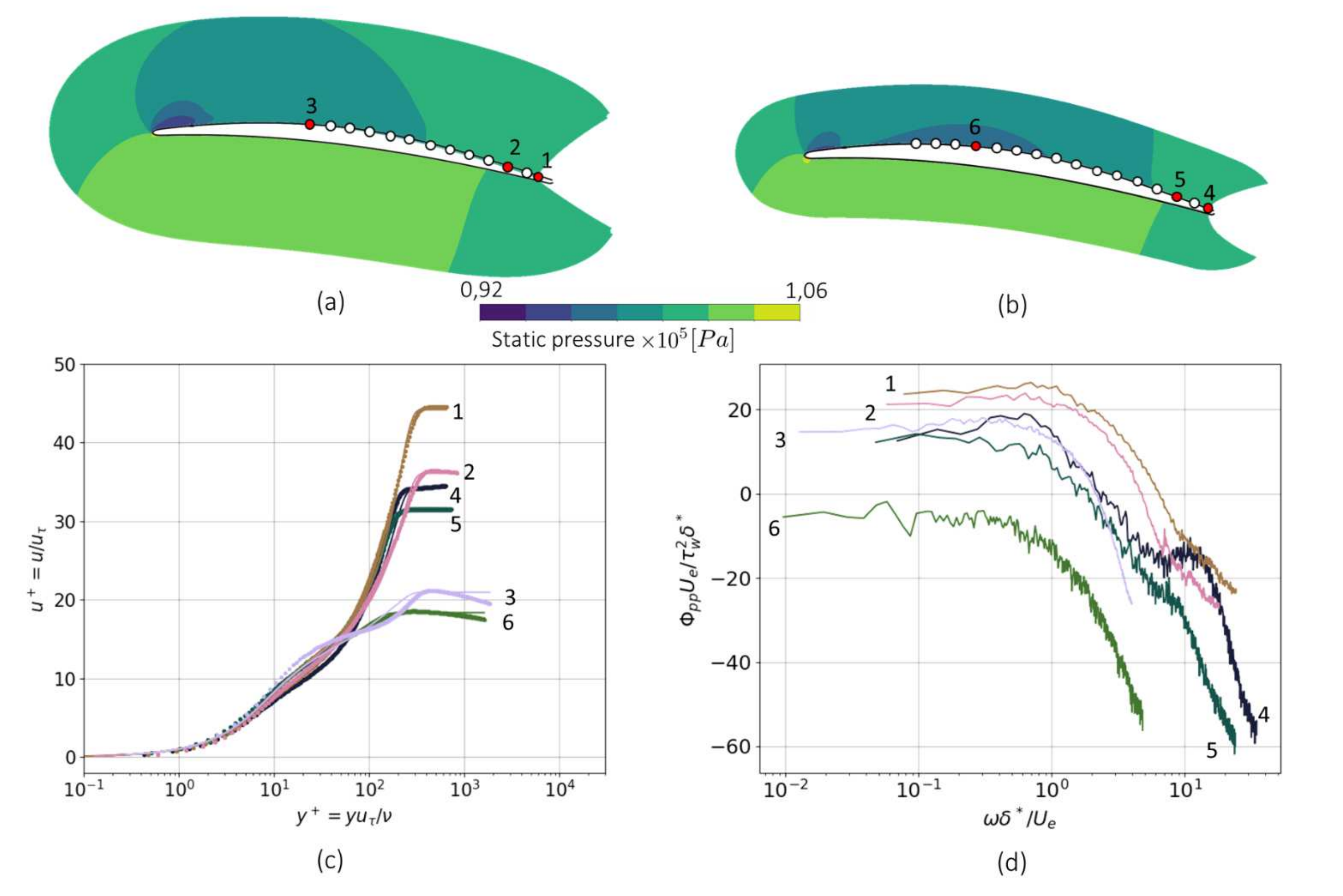}
\caption{Comparison of the numerical wall pressure spectra and velocity profile extracted perpendicular to the airfoil from (a) Deuse and (b) Hao. The turbulent boundary layer velocity profiles are expressed in wall units where the dotted points correspond to the numerical points and the lines are the fit to the model in Eq.\ref{eq:universal_vel} using the methodology in Sec. \ref{sec:5}}
\label{fig:cd_comp}
\end{figure*}

Deuse\cite{deuse2019} performed a DNS simulation of the CD airfoil using the finite difference solver named HipSTAR. The airfoil is embedded in an infinite uniform flow at a free-stream Mach $Ma=0.2$ with an angle of attack of $\alpha_e = 8^{\circ}$ and a Reynolds number $Re = 10^5$. The contour of the time-averaged static pressure field for this test case is shown in Fig.~\ref{fig:cd_comp}(a), together with the location at which the boundary layer profiles and the wall pressure spectra are sampled in this work. 

The first point, from the trailing edge, is located at 2\% of the chord upstream of the airfoil trailing edge. From there, the profiles are taken at steps of 5\% chord lengths until 60\% of the chord. Points situated further upstream are not suited for the modelling of attached TBL because the flow separates.

The velocity profiles and their associated wall pressure spectra at three selected locations (labelled with $1,2,3$ in Fig.~\ref{fig:cd_comp}(a)) are shown in Fig.~\ref{fig:cd_comp}(c) and Fig.~\ref{fig:cd_comp}(d) respectively.  The spectrum with the largest overall amplitude corresponds to the closest location to the trailing edge, and the other spectra decrease monotonously as the extraction location moves upstream. The fast-growing adverse pressure gradient explains the sharp increase of the pressure fluctuations near the trailing edge. The spectral levels also appear to be considerably larger than those obtained by Salze (Fig.~\ref{fig:salze_data}) for more moderate adverse pressure gradients. Finally, the inertial range is not observed in this simulation because of the low Reynolds number.

\subsection{Hao's CD airfoil TBL at 4 degrees of effective angle of incidence}

Hao\cite{wu2018} has also used HipSTAR to perform a simulation similar to that of  Deuse at $Ma=0.25$,  and a Reynolds number of $1.5\times 10^5$. The sampling locations for the wall pressure fluctuations and velocity profiles are the same as in the previous section for Deuse's dataset but include three extra points until 75\% of the chord. The sampling locations are shown in Fig.~\ref{fig:cd_comp}(b) together with the time-averaged static pressure field. Three selected averaged velocity and wall pressure profiles (labelled as $4,5,6$ in Fig.~\ref{fig:cd_comp}(b)) are shown in Fig.~\ref{fig:cd_comp}(c) and Fig.~\ref{fig:cd_comp}(d).

Differently from Deuse\cite{deuse2019}, Hao simulated the finite dimensions of the wind tunnel jet in which the airfoil was placed in the experiments \cite{Padois2016}. Since a lifting airfoil placed in a jet induces a deflection of the stream, the practical angle of incidence of the airfoil is reduced compared to the infinite uniform flow case. An effective angle of incidence can however be estimated, which was found to be $\alpha_e = 4^{\circ}$ in this case. As a result, the aerodynamic loading around the airfoil is reduced compared to that of Deuse\cite{deuse2019}. This explains the lower spectral levels obtained by Hao\cite{wu2018} as visible in Fig.~\ref{fig:cd_comp}(d). Another noticeable difference is the high-frequency hump observed for large pressure gradients. The precise mechanism leading to this high-frequency hump remains unclear, but we conjecture that it may be caused by the nearby turbulence over the pressure side or by the near wake.

\subsection{Christophe's CD airfoil TBL at effective angles of incidence varying between 2 and 6 degrees}

Christophe\cite{christophe2009} performed six LES simulations on the CD airfoil at a free-stream Mach number of 0.05, and effective angles of incidence $\alpha_e $ varying between $2^{\circ}$ and $6^{\circ}$. 

The velocity and the wall pressure spectra profiles are extracted from 5\% until 65\%, with steps of 5\%, of the chord upstream the airfoil trailing edge. This dataset, initially developed for uncertainty quantification, yields different boundary layer transitions and histories, affected by the presence and extent of the laminar separation bubble at the suction side. The wall pressure spectra were deemed reliable up to a frequency of about 4\,kHz, beyond which numerical dissipation caused an artificial attenuation of the pressure levels. 

\begin{table*}[htbp]
\begin{center}
\includegraphics[width=\textwidth]{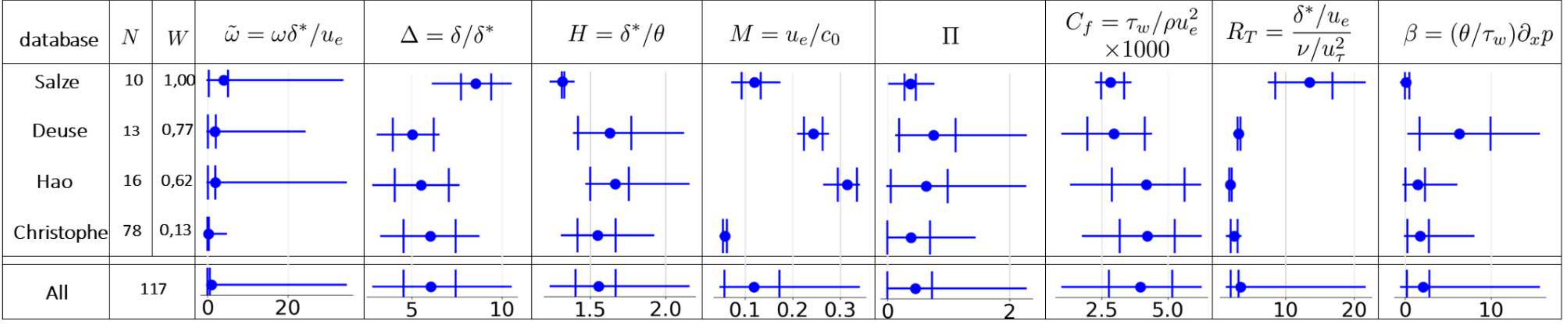}
\end{center}
\caption{Range of turbulent boundary layers parameters covered by the four datasets. The points represent the mean value while the vertical lines represent respectively the 25\% and 75\% percentiles.\vspace*{-3ex}}
\label{fig:inputrange}
\end{table*}

\subsection{A Note on the full dataset}\label{Note}

The range of dimensionless numbers covered by the datasets is illustrated in Tab.~\ref{fig:inputrange} (Table \ref{tab:input} in appendix provides the exact values and Section \ref{sec:5} describes the methodology used in their computation). 

The number of boundary layer profiles (and corresponding wall pressure spectra) for each test case is denoted as $N$ in column 2. Because these datasets have vastly different sizes, their relative contribution to the training of the ANN is weighted by the weights $W$ in column 3 (see section \ref{sec:subAAN}). This avoids over-representing some flow conditions or geometries that are more densely sampled. For the same reason, since the frequency resolution is not the same for all profiles, all wall pressure spectra have been re-sampled over 500 points logarithmically spaced over the frequency axis. 

The combination of experimental and numerical data covers a wide range of operating conditions. For example, the experimental data cover higher Reynolds numbers $R_T$ cases than the numerical simulations of the CD airfoil, but the latter offer a broader range of pressure gradients, with associated Clauser parameter $\beta$ and wake parameter $\Pi$.    

Figure.~\ref{fig:cd_comp}(d) reveals that fairly similar non-dimensional wall pressure spectra can be obtained in largely different flow conditions, at least over the analyzed frequency range. For example, the wall spectrum obtained by Hao close to the trailing edge (position ~4) of the airfoil closely matches the one sampled by Deuse at 60\% of chordwise distance from the trailing edge (position ~3), up to $\omega \delta^*/U_e=2$. Yet, the corresponding boundary layer profiles (Fig.~\ref{fig:cd_comp}(c)) are significantly different. Conversely, the available dataset doesn't contain samples with identical boundary layer parameters leading to  different wall pressure spectra. This would make the problem of identifying models in the form of \eqref{Model_Phi}  ill-posed, since the modelling would require additional parameters. 

It is nevertheless possible to give a qualitative overview of the complexity of the function to be derived by mapping the full input space sampled in the dataset onto a plane. This exercise in high-dimensional cartography can be carried out using a dimensionality reduction approach known as t-SNE (t-Distributed Stochastic Neighbor Embedding\cite{vanDerMaaten2008}). Like many other nonlinear dimensionality reduction techniques (see Maaten \emph{et al}\cite{Maaten2009DimensionalityRA} for a review), the t-SNE maps a large dimensional dataset into a lower dimensional one while preserving some nonlinear metrics of similarity. In other words, points that are close or far away in the initial space (according to a certain metrics) will also be close or far away (according to another metrics) in the reduced space provided by the t-SNE embedding.

In the t-SNE, the similarity metrics is taken using Gaussian (in the higher dimensional space) and t-Student distributions (in the low dimensional space). More specifically, given $\boldsymbol{x}'_i$ and $\boldsymbol{x}'_j$ two of the $n_p$ vectors in the original space $\mathbb{R}^{n_x}$, the degree of similarity under a Gaussian distribution is 

\begin{equation}
\label{d_ij_EQ}
q_{i,j}=q(\boldsymbol{x}'_i,\boldsymbol{x}'_j)=\frac{\exp(-||\boldsymbol{x}_i-\boldsymbol{x}'_j||^2/(2\sigma^2_i))}{\sum_{k\neq j} \exp(-||\boldsymbol{x}'_i-\boldsymbol{x}'_j||^2/(2\sigma^2_i))}\,.
\end{equation}\, where $\sigma_i$ is a user defined parameter and assuming that both inputs have been normalized in the range $[0,1]$. Hence two samples which are similar (close to each other in $\mathbb{R}^{n_x}$) leads to large $q_{i,j}$ while samples that are largely different (far away in $\mathbb{R}^{n_x}$) leads to $q_{i,j}\approx 0$. The normalization in \eqref{d_ij_EQ} allows to interpret $q_{i,j}$ as a distribution. A similar metric $d_{i,j}(\boldsymbol{z}_i,\boldsymbol{z}_j)$ is defined in the reduced space (albeit with a different distribution) and the mapping can be done in such a way that the pair-wise distances are preserved as much as possible.

\begin{figure}[h]
    \centering
    \includegraphics[width=0.5\textwidth]{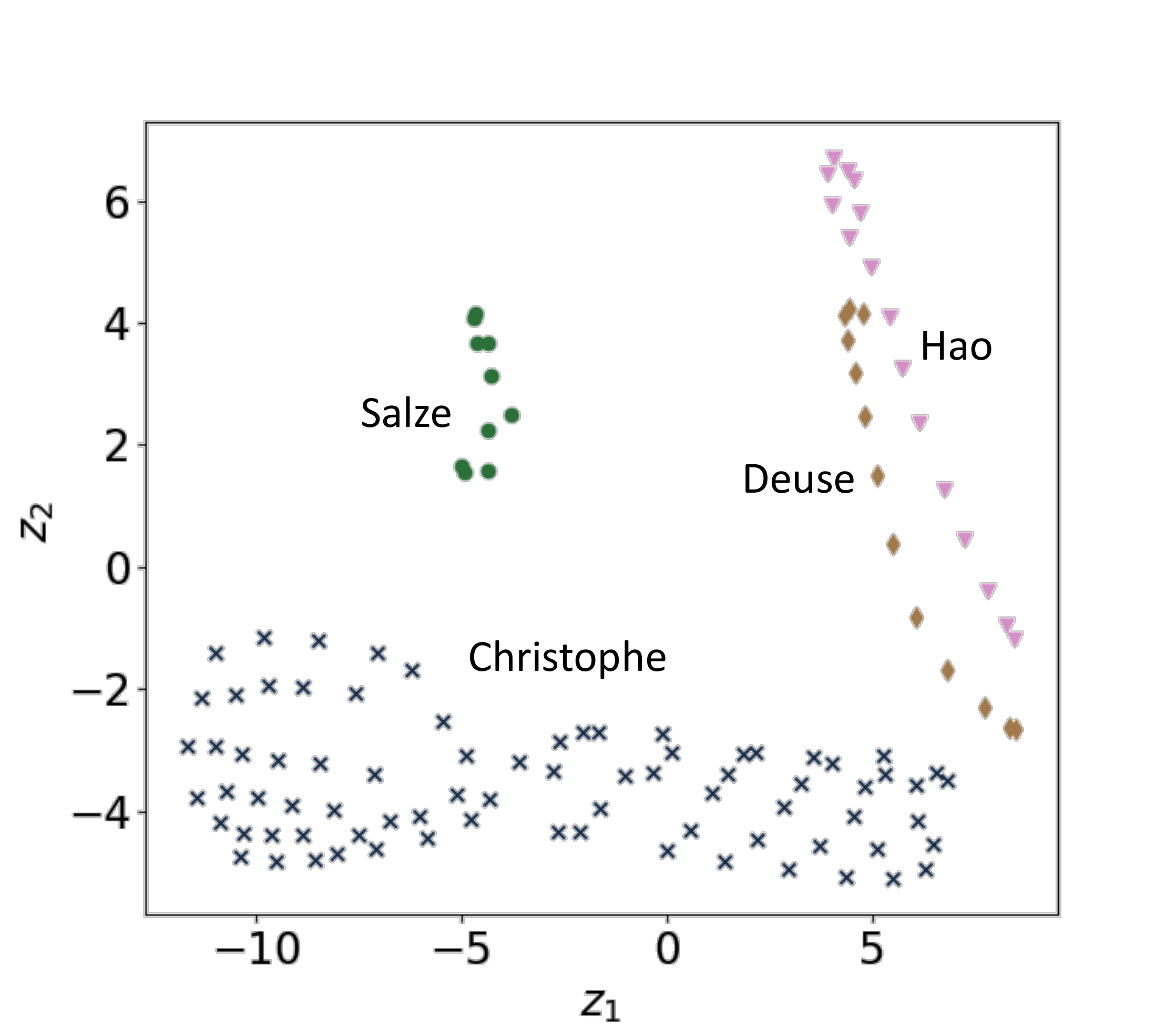}
         \caption{T-SNE manifold 2D map of the 7D inputs of Tab. \ref{fig:inputrange}. \textcolor{color11}{\ding{53}} for Christophe, \textcolor{color12}{\ding{108}} for Salze, \textcolor{color13}{\ding{117}} for Deuse and \textcolor{color14}{\ding{116}} for Hao.}
    \label{fig:tsne}
\end{figure}

\begin{figure}[h]
    \centering
    \includegraphics[width=0.5\textwidth]{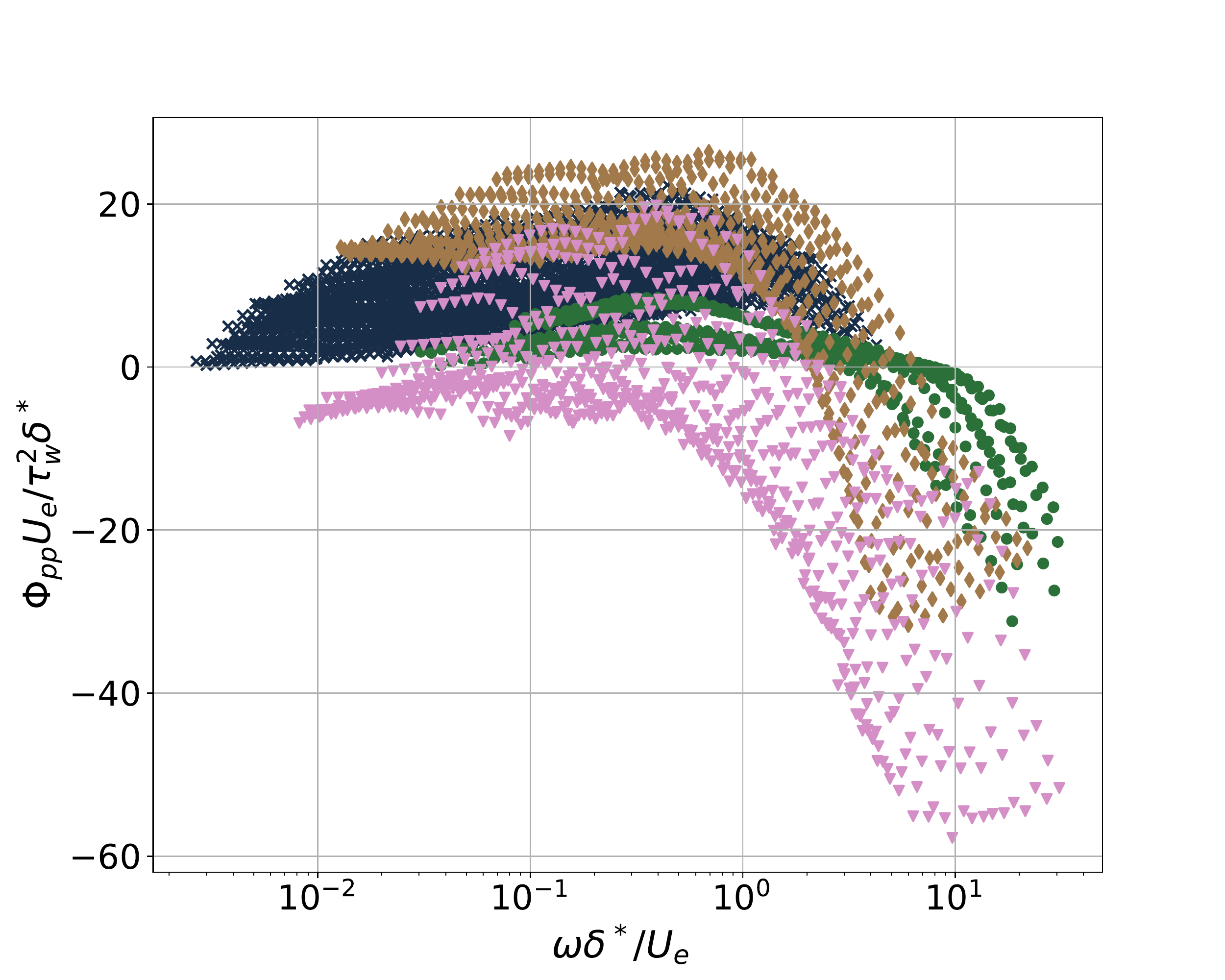}
         \caption{Wall pressure spectra \textcolor{color11}{\ding{53}} for Christophe, \textcolor{color12}{\ding{108}} for Salze, \textcolor{color13}{\ding{117}} for Deuse and \textcolor{color14}{\ding{116}} for Hao.}
        \label{fig:tsne2}
\end{figure}

We use this approach to visualize the input dataset. For a given wall pressure spectra, we map the input space for the wall pressure spectra model, i.e. $\boldsymbol{x}'=[\Delta, H, Ma, \Pi,C_f,R_T, \beta]\in\mathbb{R}^{7}$, into a two-dimensional plane $\boldsymbol{z}=({z}_1,{z}_2)$. The result is shown in Fig. \ref{fig:tsne}, with different markers denoting the datasets previously introduced. 
The figure shows that Deuse's and Hao's datasets are close (according to a metric like \eqref{d_ij_EQ}) in the input space while Salze's is different from all the others. Christophe's dataset covers the largest span and has conditions similar to Deuse's (at high $z_1$). This figure should be analyzed together with Fig. \ref{fig:tsne2}, collecting all the 117 available power spectral densities, labelled by the dataset with the same markers as in Figure \ref{fig:tsne}. 

The plot confirms the expectation that the wall pressure spectra in Hao and Deuse are relatively similar, given their similarity in the input space. However, their spectral amplitudes vary significantly, potentially indicating strong sensitivity to the inputs. The amplitude of the wall pressure spectra in Salze falls somewhere in between the one of Hao and Deuse despite the largely different input space. However, Salze's wall pressure spectra differ significantly in shape since the inertial range is clearly observable contrary to the spectra obtained on the CD airfoil.

While the available dataset does not allow for assessing the well-posedness of the regression problem, it is clear that the function at hand is particularly complicated. As we could not find similar turbulent boundary layers with different wall pressure spectra, we have no evidence that the development history of the non-equilibrium wall pressure spectra used in this study is not already accounted for by the set of parameters from \eqref{eq:Buckingham}.

\section{Methodology}
\label{sec:5}

\subsection{Computation of the boundary layer parameters}\label{sec:5p1}

\begin{figure}[h!]
\includegraphics[width=0.4\textwidth]{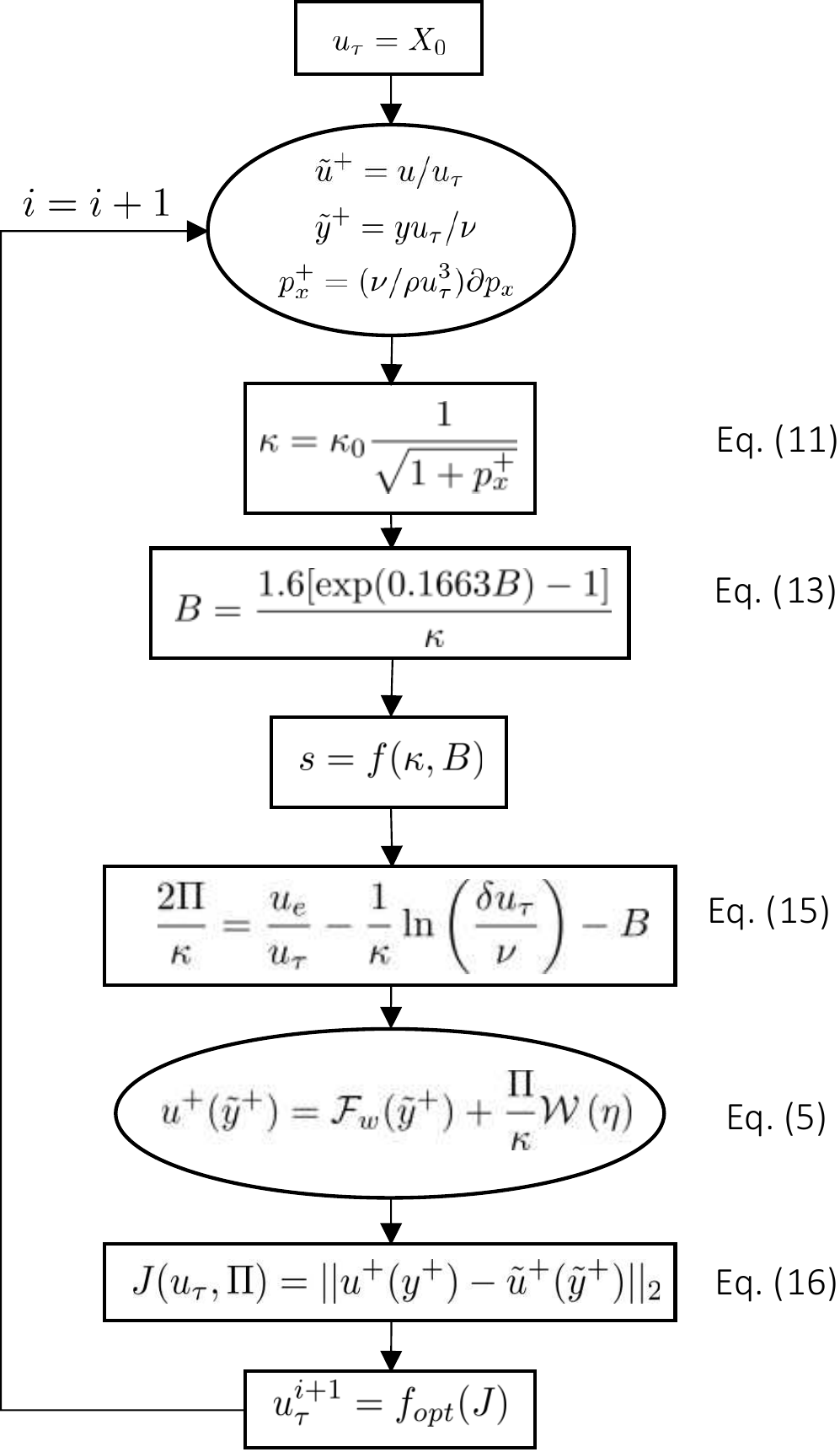}
\caption{Computation loop for the error used within the optimisation of the friction velocity $u_\tau$ and wake parameter $\Pi$.}
\label{fig:bl_model}
\end{figure}

The boundary layer parameters $\Pi$ and $\tau_w$ are obtained from a nonlinear regression of the boundary layer velocity profile $u(y)$ by the model laid out in Eq.~(\ref{eq:universal_vel}). The method is inspired by Clauser's method \cite{clauser1956turbulent} and the work of Kendal \cite{kendall2008} and Rodríguez‑López \cite{rodriguez2015} for the determination of the wall shear stress by fitting the data to a theoretical velocity profile. We extend this approach with an iterative approach, illustrated in Fig.~\ref{fig:bl_model}, to account for the presence of APG or FPG. While Clauser's method adjusts the friction velocity $u_\tau$ to match the universal logarithmic law in the overlap region of the boundary layer, we adjust both $u_\tau$ and the wake parameter $\Pi$ to fit the extracted data with a complete velocity profile accounting for the presence of a pressure gradient.

Following Eq.~(\ref{eq:universal_vel}) in Section \ref{sec:2}, the law of the wall $\mathcal{F}_w(y^+)$ used in this work is the one proposed by Musker \cite{musker}:
\begin{equation}\label{eq:musker}
    \mathcal{F}_w(y^+) = \int_0^{y^+}\frac{s \xi^2 + \kappa }{\kappa \, s\, \xi^3+ s\,\xi^2+\kappa} d \xi,
\end{equation}
with $\xi$ a dummy integration variable, $\kappa$ the von Karman constant and $s$ the free parameters of this formulation. For a ZPG, setting $s=0.001093$ and $\kappa=0.41$ results in a velocity profile matching the classic logarithmic profile $1/\kappa \,ln(y^+)+B$ with $B=5.0$. This formulation has the main advantage of not requiring a piece-wise definition of the law of the wall, which is also usually problematic in the buffer region ($5\leq  y^+ \leq 30$).

To account for the pressure gradient effect, we follow Nickels' approach\cite{nickels} and modify the von Karman constant as:
\begin{equation}
\label{eq:nickel_kappa}
\frac{\kappa}{\kappa_0} = \frac{1}{\sqrt{1 + p_x^+ y_c^+}}  \,
\end{equation} where $\kappa_0 = 0.41$ is the von Karman constant for ZPG, and $y_c^+$ is the point at which the logarithm layer intersects the linear prediction in the viscous sub-layer. This quantity is also a function of the pressure gradient, and can be computed as the smallest positive root of the cubic:
\begin{equation}
\label{eq:nickel_yc}
p_x^+(y_c^+)^3 + (y_c^+)^2 - 144 = 0,   
\end{equation} where $p^+_x= (\delta_\nu/\theta)\beta$ is the dimensionless pressure gradient. 
Combining \eqref{eq:nickel_yc}-\eqref{eq:nickel_kappa}, one sees that $\kappa > \kappa_0$ for APG and $\kappa < \kappa_0 $ for FPG. This is consistent with what observed in this study and with the discussion on the non-universality of the von Karman coefficient by Nagib \emph{et al} \cite{nagib}. These authors have also introduced an empirical relation to link $\kappa$ and $B$ in the logarithmic region ($y^+\geq 30$) in the presence of a pressure gradient:
\begin{equation}
\label{eq:nagib}
B  = \frac{1.6[\exp(0.1663B)-1]}{\kappa} \,.
\end{equation}

The alteration of the von Karman constant permits a better fit in the logarithmic region when a pressure gradient is present, and also influences indirectly the wake parameter $\Pi$.

The wake function $\mathcal{W}(\eta)$ in Eq.~(\ref{eq:universal_vel}) used in this work is the one proposed by Chauhan \cite{chauhan2007}: 
{\fontsize{8}{8}\selectfont %
\begin{align}
\label{Chauhan}
\mathcal{W}(\eta) &=\frac{1 -\exp\left(-0.25\left(5a_2 + 6a_3 +7a_4)\eta^4 + a_2\eta^5 + a_3 \eta^6 + a_4\eta^7\right)\right)}{1-\exp(-0.25(a_2+2a_3+3a_4))} \nonumber  \\
  & \times \left( 1-\frac{1}{2\Pi} \ln(\eta) \right),
\end{align}}
where $a_2 = 132.84$, $a_3 = -166.2$, $a_4 = 71.91$ ensure a continuous slope at the edge of the wake region. The wake parameter can be identified by noticing that the normalizing condition imposes $\mathcal{W}(1)=2$ and thus at the edge of the boundary layer one has
\begin{equation}
    \label{eq:wake}
    \frac{u_e}{u_\tau} = \frac{1}{\kappa} \ln \left(\frac{\delta u_\tau}{\nu}\right) + B+ \frac{2 \Pi}{\kappa}\,.
\end{equation}

Combining the equations introduced thus far (cf. Figure \ref{fig:bl_model}), it is possible to set up a regression problem in which the friction velocity $u_\tau$ and the wake parameter $\Pi$ are the ones that allows minimizing the loss function
\begin{equation}
{J}(u_\tau,\Pi)=||u^+(y^+)-\tilde{u}^+({y}^+)||_2
\end{equation}
where $||\bullet||_2$ is the $l_2$ error, $u^+(y^+)$ is the available data and $\tilde{u}^+({y}^+)$ is the model prediction.

 \begin{figure}[h]
\includegraphics[width=0.45\textwidth]{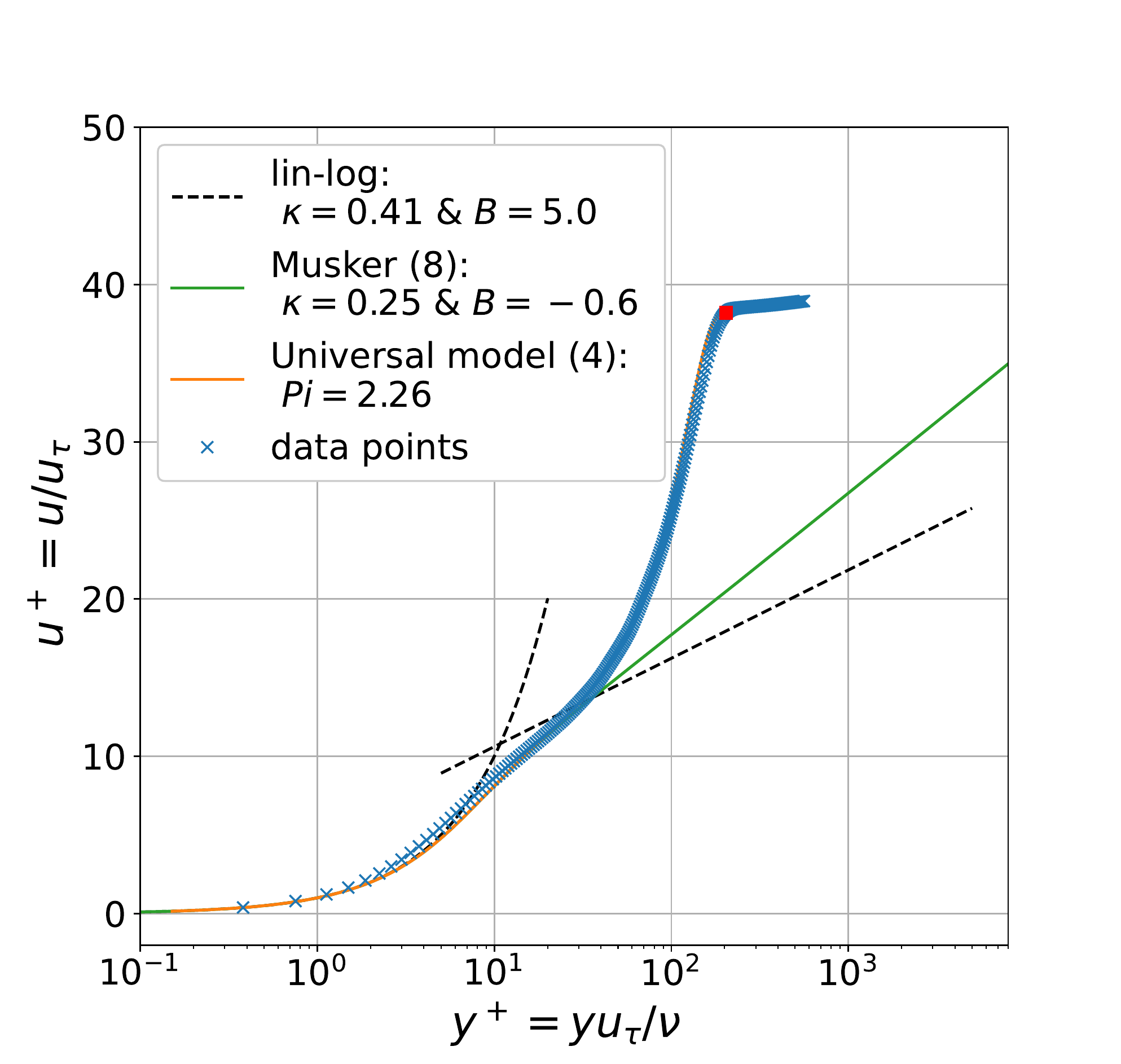}
\caption{Example of boundary layer velocity fitting with pressure gradient correction}
\label{fig:p_corr}
\end{figure}

An example of boundary layer velocity profile in presence of pressure gradient is shown in Fig. \ref{fig:p_corr}, considering a test case with a strong pressure gradient ($\beta=5.93$) from Hao at 2\% chord upstream the the airfoil trailing edge (position no~4 in Fig. \ref{fig:cd_comp}). The proposed optimization loop leads to $u_\tau=2.37$ m/s and $\Pi=2.26$, producing an excellent match with the available data. For completeness, the figure also reports the prediction of a classic piece-wise linear-logarithmic model and Musker's model in \eqref{eq:musker} with $\kappa = 0.25$ and $B=-0.6$. The classic piece-wise linear-logarithmic model is unable to match the correct slope in the logarithmic region while the Musker law modified with the pressure gradient correction from \eqref{eq:nagib} and \eqref{eq:nickel_kappa} provides an excellent match with the data.

\subsection{The Artifical Neural Network (ANN)} \label{sec:subAAN}

\begin{figure}[h!]
\includegraphics[width=0.4\textwidth]{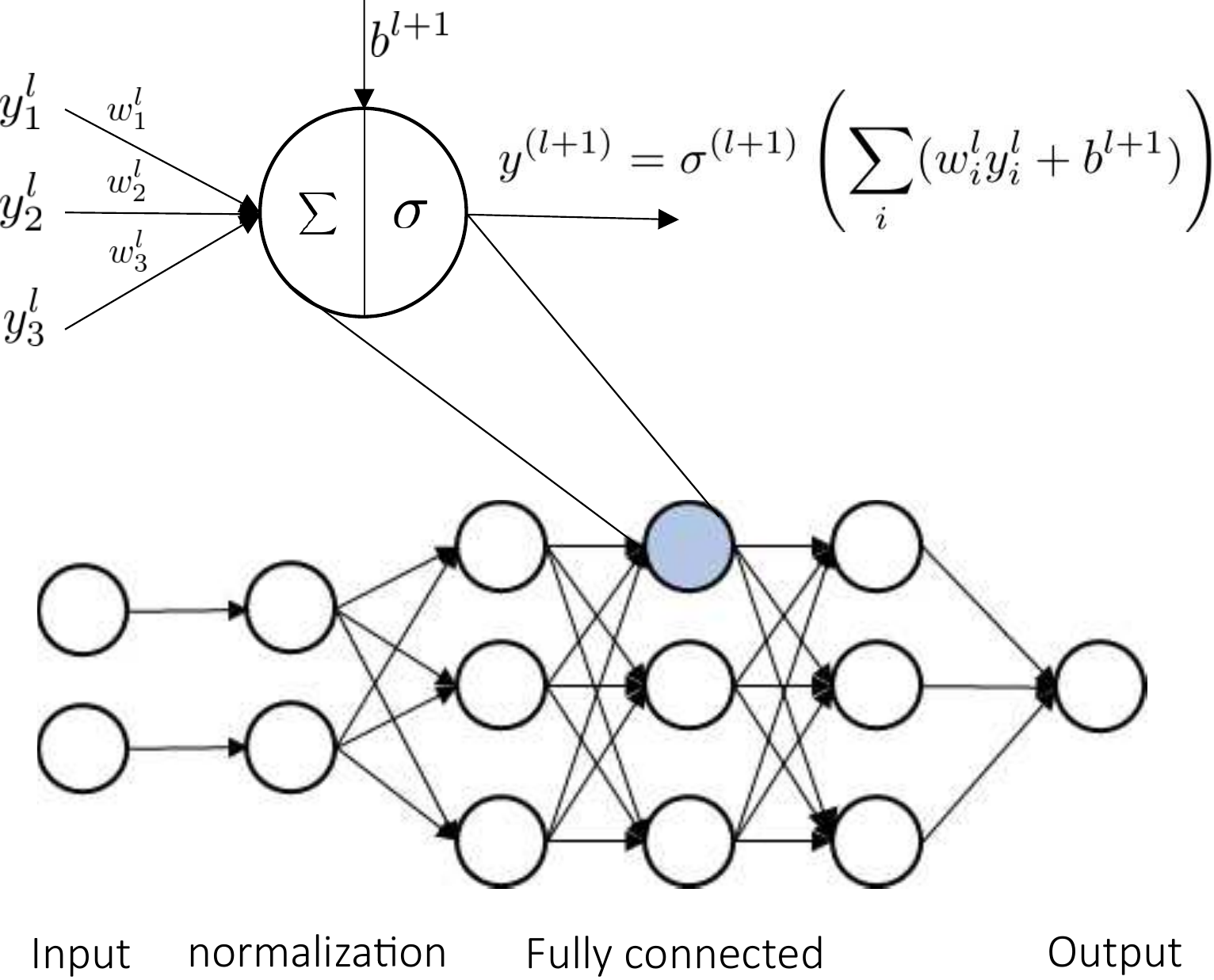}
\caption{Example of Neural network architecture and activation function.}
\label{fig:NN_architecture}
\end{figure}

ANNs are distributed architecture of simple computational units\cite{Goodfellow}, called \emph{neurons}, connected in layers as shown in Fig. \ref{fig:NN_architecture}. These architectures provide a parametric function $\boldsymbol{y}=f(\boldsymbol{x},\boldsymbol{w},\boldsymbol{b})$ mapping an input $\boldsymbol{x}\in\mathbb{R}^{n_x}$ to an output $\boldsymbol{y}\in\mathbb{R}^{n_y}$ according to a set of parameters called \emph{weights} $\boldsymbol{w}\in\mathbb{R}^{n_w}$ and \emph{biases} $\boldsymbol{b}\in\mathbb{R}^{n_b}$. In the simplest configuration used in this work, namely a fully connected feed-forward network, this function is defined in a recursive way, such that the output of each layer is the input of the following. Specifically, given $\boldsymbol{y}^{(l)}\in\mathbb{R}^{n_l}$ the output of a layer containing $n_l$ neurons, the output of the following layer is

\begin{equation}
    \label{ANN_EQ}
    \boldsymbol{y}^{(l+1)}=\sigma^{(l+1)}(\boldsymbol{z}^{(l+1)}) \,,
\end{equation} with 

\begin{equation}
    \label{ANN_EQ_2}
    \boldsymbol{z}^{(l+1)}=\boldsymbol{W}^{(l)}\boldsymbol{y}^{(l)}+\boldsymbol{b}^{(l+1)}\,,
\end{equation} and $\sigma^{(l+1)}()$ the \emph{activation} function at layer $(l+1)$, $\boldsymbol{W}^{(l)}\in\mathbb{R}^{n_l\times n_{l+1}}$ the matrix of \emph{weights} connecting the $n_l$ neurons in layer $l$ to the $n_{l+1}$ neurons in layer $l+1$ and $\boldsymbol{b}^{l+1}\in\mathbb{R}^{n_{l+1}}$ the vector of \emph{biases} at layer $l+1$. The input and outputs of the ANN are, respectively, the input of the first layer and the output of the last one.

ANNs are known as universal function approximators because they can approximate any function if a sufficiently large number of neurons and layers is used \cite{Cybenko1989}. In this work, the ANN is called to approximate eq. \eqref{eq:Buckingham}, hence the input vector is $\mathbf{x}=[\tilde{\omega},\Delta, H, Ma, \Pi, C_f, R_T, \beta ]\in\mathbb{R}^{8}$ and the output is the scalar $y=10\log_{10}(\Phi_{pp} u_e / \tau_w^2 \delta^*)\in\mathbb{R}$ which expresses the dimensionless wall pressure spectra in Decibel per Hertz. The training of the ANN consists in determining the set of weights $\boldsymbol{w}=(\boldsymbol{W}^{(1)},\cdots \boldsymbol{W}^{(n_L)})$ and biases $\boldsymbol{b}=(\boldsymbol{b}^{(1)},\cdots \boldsymbol{b}^{(n_L)})$, with $n_L$ the total number of layers, such that a loss function measuring the error in the prediction over the available data is minimized. In this work, the loss function is taken as the weighted mean square error on the logarithmic values (lMSE):

\begin{equation} \label{eq:MSE}
 lMSE(\boldsymbol{w},\boldsymbol{b}) = \frac{1}{N}\sum_{i=1}^N W_i  \left(y_i - y_{i}^{A}(\boldsymbol{w},\boldsymbol{b}) \right)^2
\end{equation} where $ y_{i}^{A}(\boldsymbol{w},\boldsymbol{b})$ is the prediction of the ANN for the $i-th$ sample, $y_i$ is the $i-th$ available data point and $W_i$ is the weight (shown also in table \ref{fig:inputrange}) of each data point. These weights are chosen to give equal importance to the four available datasets despite their widely different sizes. In the machine learning literature, this is a common approach when dealing with datasets in which some classes or some portions of the sample space is more present than others in the training data\cite{cui2019}.

The training of the network was carried out using Nadam optimizer \cite{dozat2016} with a learning rate of $10^{-4}$. This is a variant of the stochastic gradient descent which uses inertia during the descent and the classic back-propagation algorithm\cite{David} to compute the gradient of the loss function with respect to the ANN parameters. Following a mini-batch approach, the gradient of the loss function is computed on a batch of $n_B=32$ randomly chosen training sample.

During the training, the dataset is split into training data ($80\%$) and validation data ($20\%$). The lMSE in the first drives the optimization (training), while the lMSE in the second allows for identifying the stopping point using the early stopping approach. This stops the training when the validation error stops decreasing within a user-defined tolerance. In addition, six profiles were removed from the dataset and used as testing data to assess the ANN's ability to generalize beyond the training data. The performances of the ANN on these profiles is illustrated in Sec. \ref{sec:6}.

\begin{figure}[h]
\includegraphics[width=0.45\textwidth]{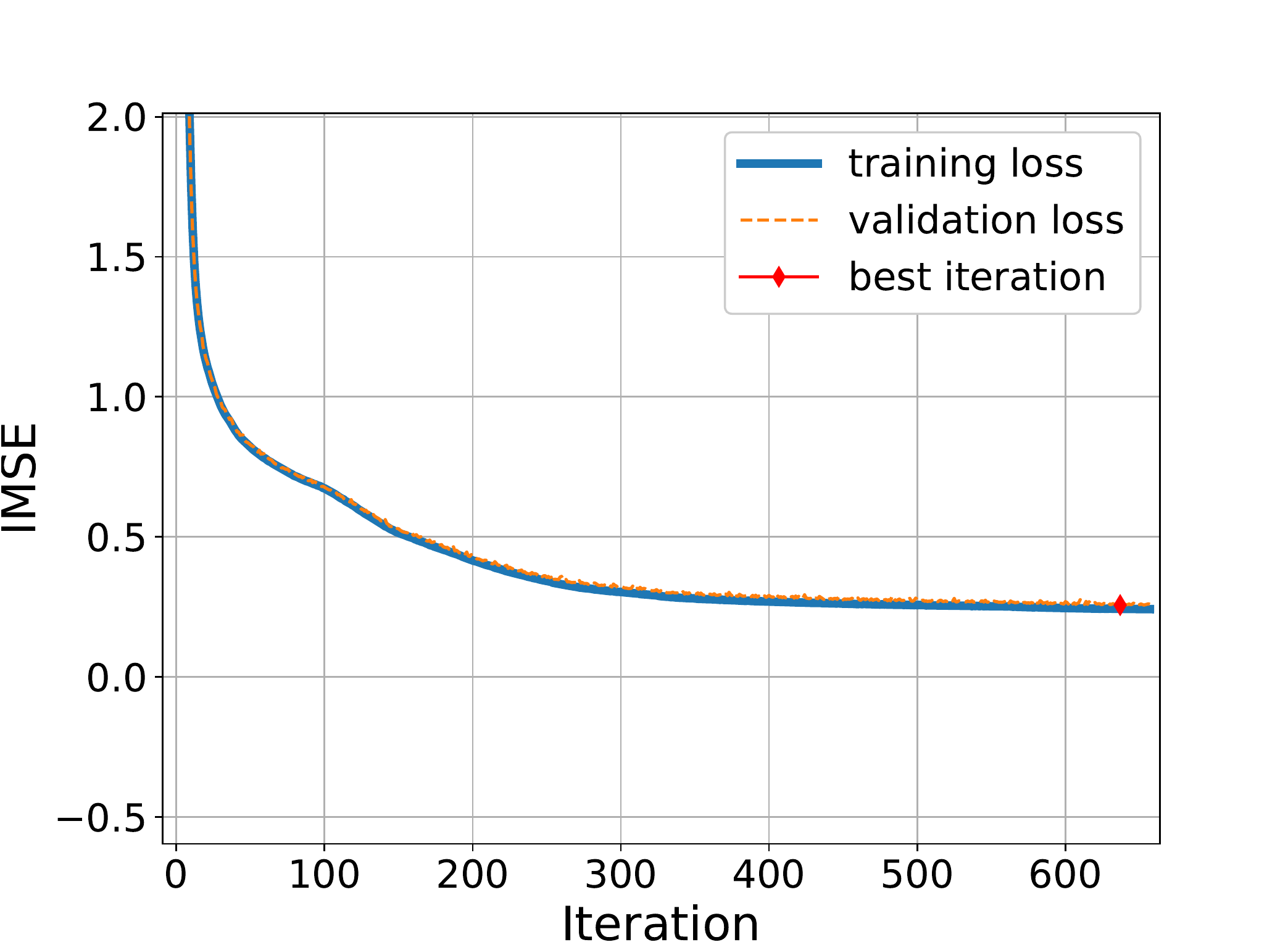}
\caption{Evolution of training and validation error.}
\label{fig:NN_training}
\end{figure}

Fig.~\ref{fig:NN_training} shows a typical learning curve for the ANNs used in this work. These have four layers between the input and outputs. The first is a normalization layer that scales the inputs using their mean and standard deviation, as this helps to stabilize and accelerate the learning process \cite{ba2016}. The remaining layers are fully connected with $n_N=10$ neurons and Selu activation functions, as recommended with normalized inputs \cite{klambauer2017}.

\begin{figure}[h]
\includegraphics[width=0.45\textwidth]{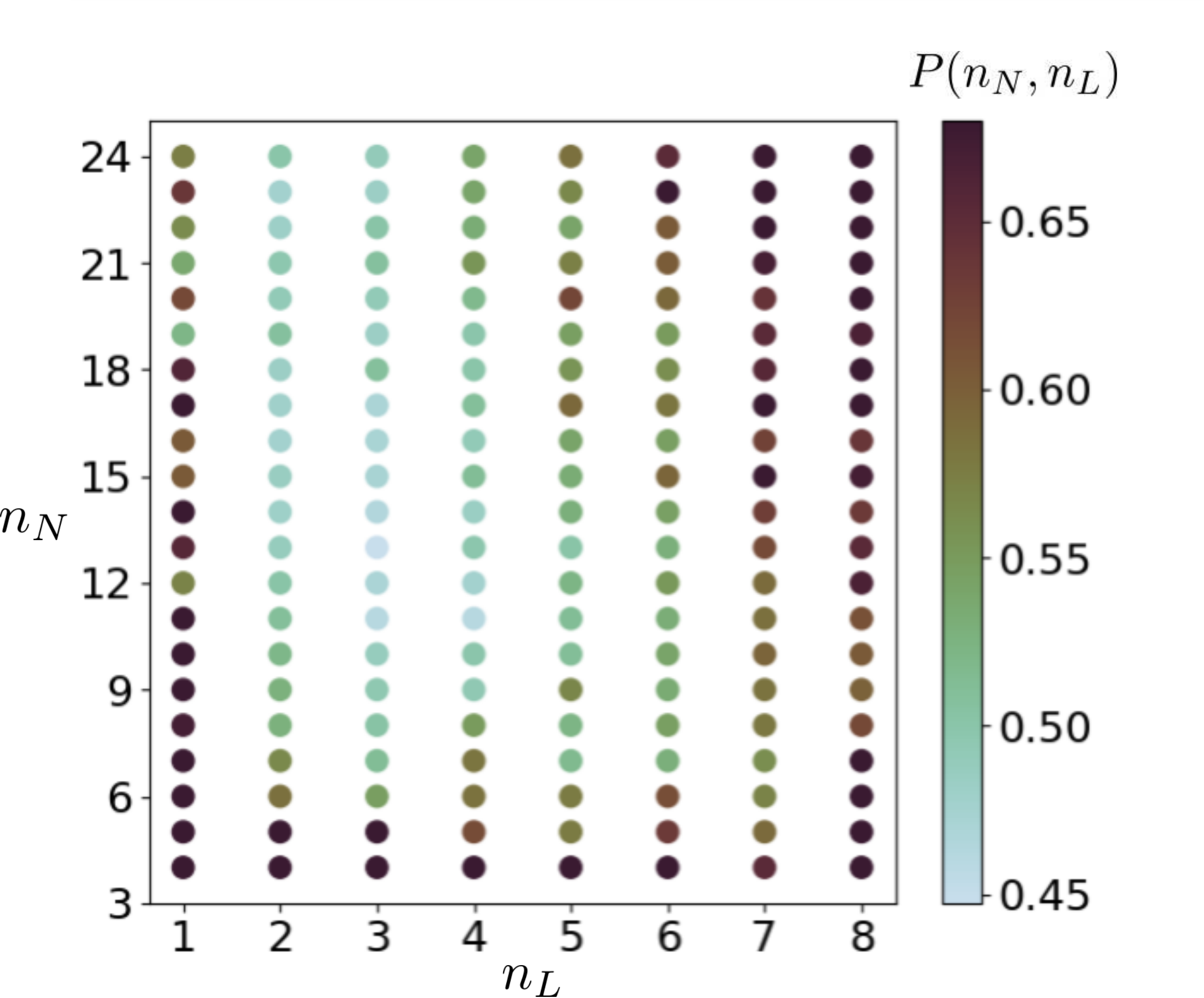}
\caption{Performance indicator \eqref{Perf} for different network architecture composed of $nL$ layers and $nN$ nodes.}
\label{fig:NN_geometry}
\end{figure}

The proposed architecture resulted as the most appropriate compromise between model complexity and accuracy. Considering layers of equal size ($n^{(1)}=n^{(2)}\dots=n_N$), we performed a sensitivity study testing different combinations of $n_N$ and $n_L$. For each combination, an ANN was trained and a performance measure was defined as 

\begin{equation}
\label{Perf}
    P(n_N,n_L)=lMSE+0.01 ||(\boldsymbol{w},\boldsymbol{b})||_F
\end{equation} where $||(\boldsymbol{w},\boldsymbol{b})||_F$ is the Frobenious norm of the weights and biases in the network. This term acts as a regularization by penalizing solutions with large weights. The analysis shows that different architectures lead to comparable performances on the available data as long as they include sufficient parameters. However, increasing the number of layers and nodes increases the computational cost and the chances of overfitting the data, causing poor generalization. Similar performances could be obtained by ANNs with either $n_L=2$ with $n_N=15$ to $24$ or $n_L=3$ layers with $n_N=10$ to $20$ (\textsc{Fig.}~\ref{fig:NN_geometry}). Among these, the option with fewer number of weight and biases was favoured, leading to $n_L=3$ and $n_N=10$. 


\subsection{The ANN uncertainties}

We estimate the uncertainties of the ANN model using an ensemble learning approach\cite{Abdar2021,Gawlikowski2021}. The focus is placed on the model (epistemic\cite{Huellermeier2021}) uncertainties, as these allow quantifying the performances of the ANN in relation to the available data. The epistemic uncertainties are linked to the uncertain determination of the ANN's weight and biases. We are here interested in their impact on the model predictions, especially in regions of the input space that have not been sampled adequately. Besides providing confidence intervals, these allow unveiling regions of the input space that lack training data.

The ensemble approach implemented in this work (see also Lee \emph{et al}\cite{Lee2015} and Lakshminarayanan \emph{et al}\cite{Lakshminarayanan2016}) consists in training an ensemble of $n_E=100$ versions of the same ANN. These networks are trained on the same dataset of $111$ profiles (having removed 6 for testing purposes), but differ because of the different (random) initialization of weight and biases and because of the inherent stochasticity of the training process. Each network is trained with a different (random) splitting between training and validation data ($80\%$ and $20\%$ respectively) and different (random) batch selection during the computation of the loss function gradients.

For a given input vector $\boldsymbol{x}=[\tilde{\omega},\Delta, H, Ma, \Pi, C_f, R_T, \beta ]\in\mathbb{R}^{8}$, given $y^{A,j}$ the prediction of the $j^{th}$ ANN, the ensemble prediction and variance are given as 
\begin{equation}\label{eq:ANN_epistemic}
   \mu_{e}(\boldsymbol{x})=\mathbb{E}_{\sim E}\{y^{A,j}\}\quad \mbox{and}\quad
    \sigma^2_{e}(\boldsymbol{x})=\mathbb{E}_{\sim E}\{(y^{A,j}-\mu_e)^2\}\,,
\end{equation} where $\mathbb{E}_{\sim E}$ is the expectation operator over the ensemble of networks. These contributions are function of the inputs and one would expect $\sigma_{e}(\boldsymbol{x})$ to be large for $\boldsymbol{x}$ in poorly sampled regions. 

As a simple model for the aleatory uncertainty, we consider a symmetric distribution centered in the ensemble mean. Therefore, assuming that the minimization of the lMSE in \eqref{eq:MSE} during the training ensures a symmetric distribution of the errors, the aleatory contribution to the uncertainty is constant for the entire input space and can be estimated as 
\begin{equation}\label{eq:ANN_aleatory}
\sigma^2_{a} = \mathbb{E}_{\sim E} \{ lMSE_j\}\,.
\end{equation} 

This estimate is optimistically biased because it assumes that the training data is rich enough to have converged statistics and assumes that the same statistics hold outside the training range. Nevertheless, this simple estimate fits the purpose of providing a first glance at the ANN predictive capabilities and succeeds in providing confidence intervals that accommodate the training, the validation, and the test data.

The standard deviation due to the aleatory contribution was found to be equal to $0.54$ dB/Hz. This is the irreducible uncertainty of the data (due to measurements or numerical uncertainties) which would, in principle, remain unvaried even if the training of the network is enriched with more profiles.

Assuming that these contributions are independent, the standard deviation in the prediction is thus $\sigma_E=\sqrt{\sigma^2_{e}+\sigma^2_{a}}$. Taking a 95\% confidence interval for the prediction, the uncertainty intervals can be taken as $\mu_E \pm 1.96 \sigma_E$.

\begin{figure*}
    \centering
    \begin{subfigure}[b]{0.33\textwidth}   
    \centering 
    \includegraphics[width=\textwidth]{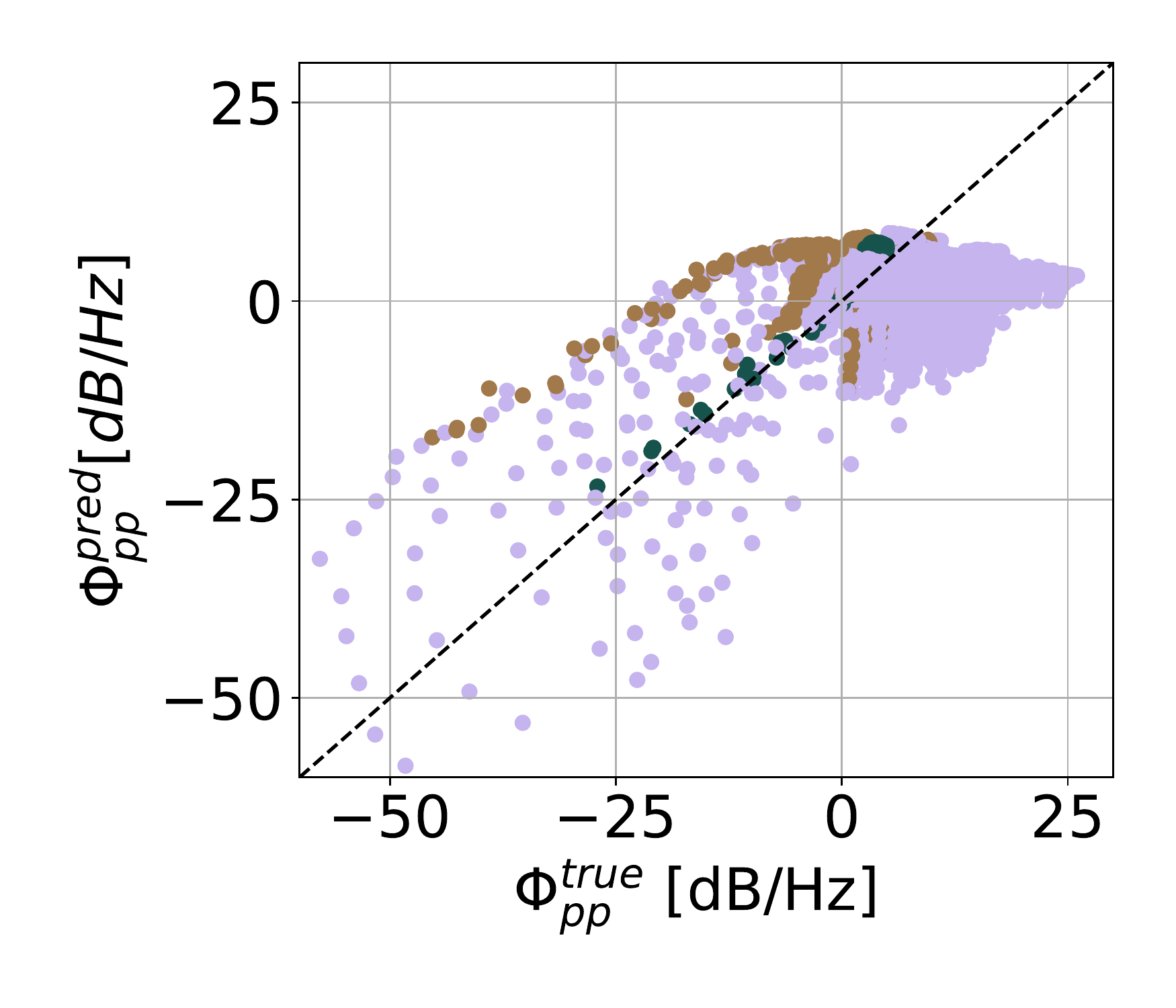}
    \caption{Goody lMSE = 90.12 dB/Hz}   
    \label{fig:perf_goody}
    \end{subfigure}
    \hfill
    \begin{subfigure}[b]{0.33\textwidth}
    \centering
    \includegraphics[width=\textwidth]{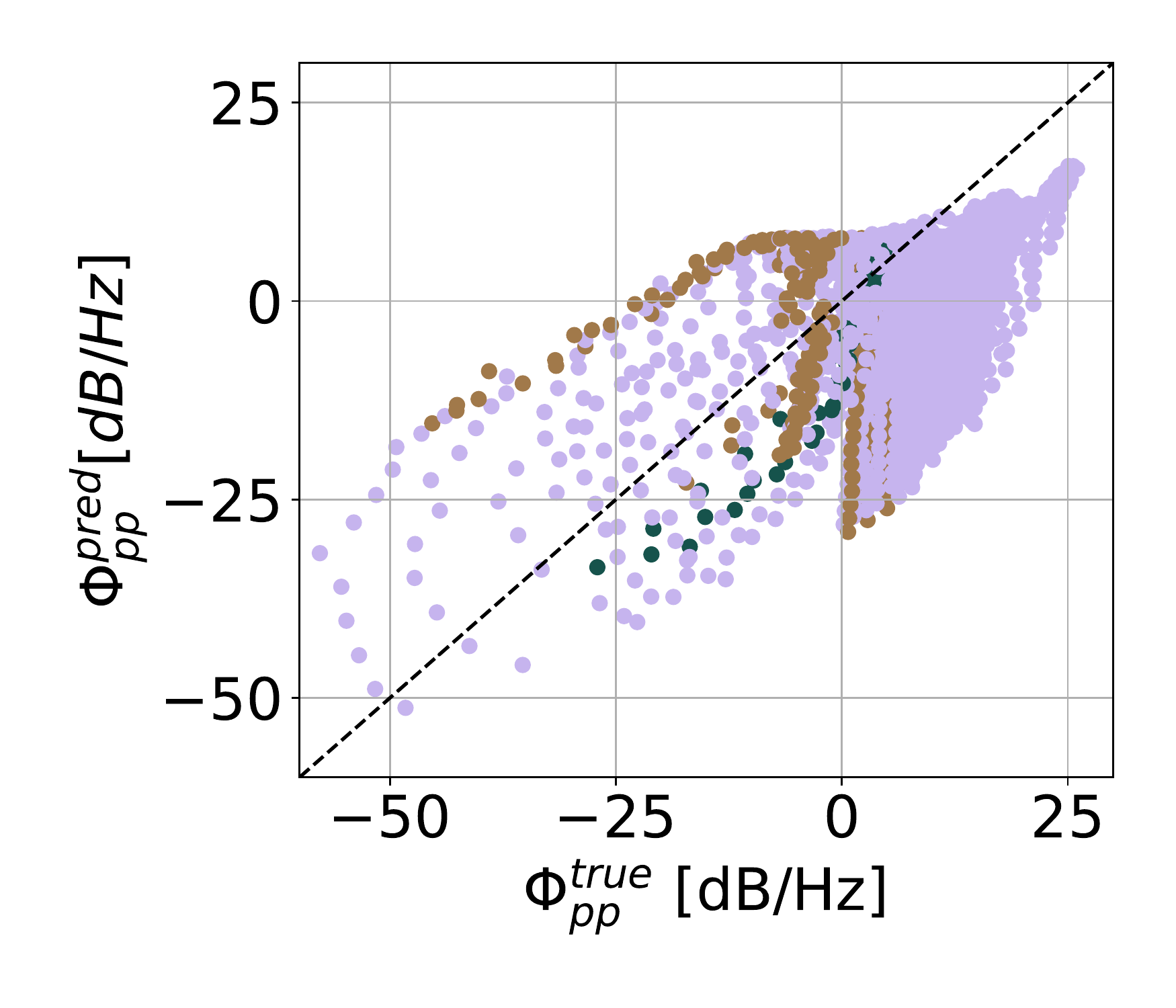}
    \caption{Kamruzzaman lMSE = 160.98 dB/Hz}
    \label{fig:perf_kamruzzaman}
    \end{subfigure}
    \hfill
    \begin{subfigure}[b]{0.33\textwidth}  
    \centering 
    \includegraphics[width=\textwidth]{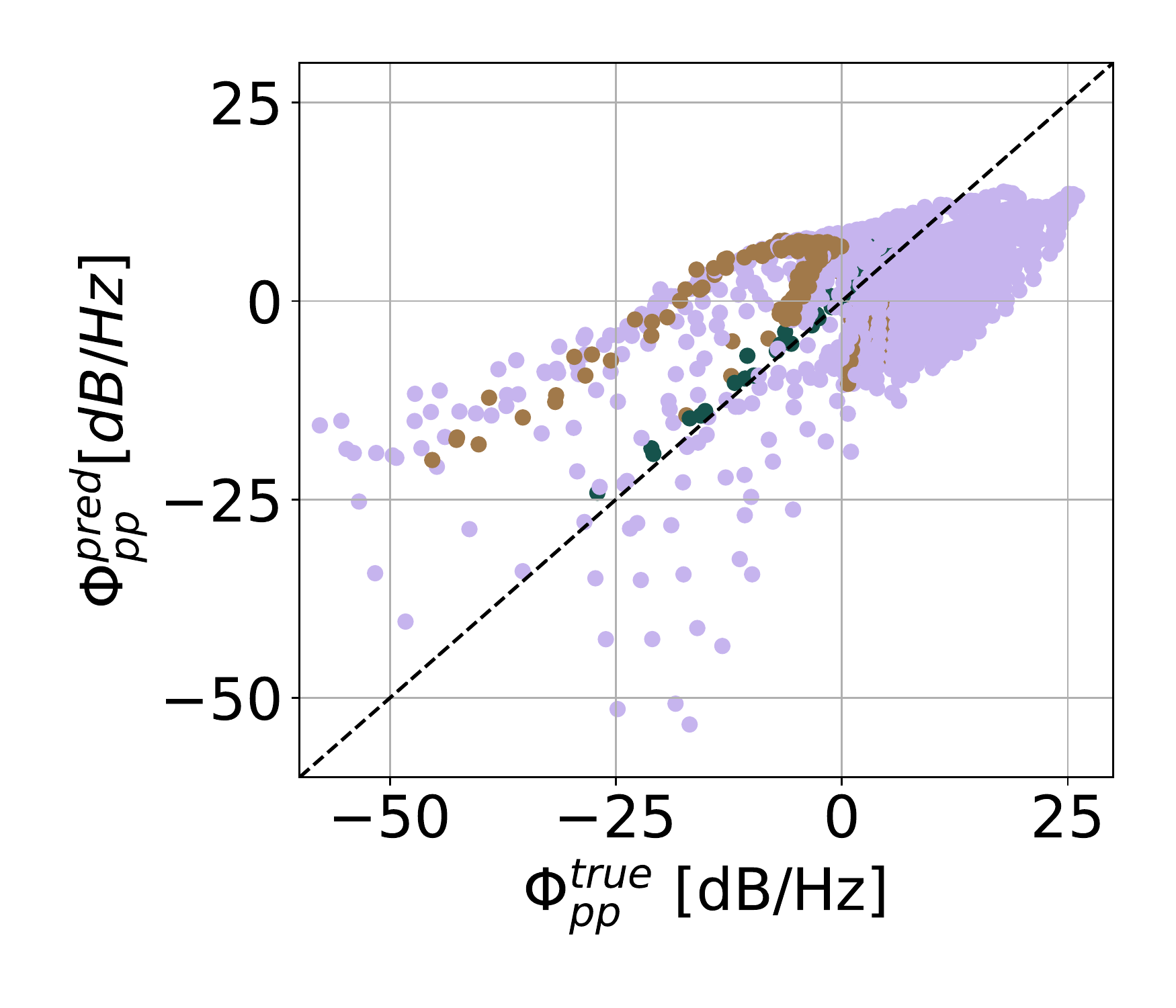}
    \caption{Rozenberg lMSE = 76.14 dB/Hz}   
    \label{fig:perf_Rozenberg}
    \end{subfigure}
    \vskip\baselineskip
\begin{subfigure}[b]{0.33\textwidth}   
    \centering 
    \includegraphics[width=\textwidth]{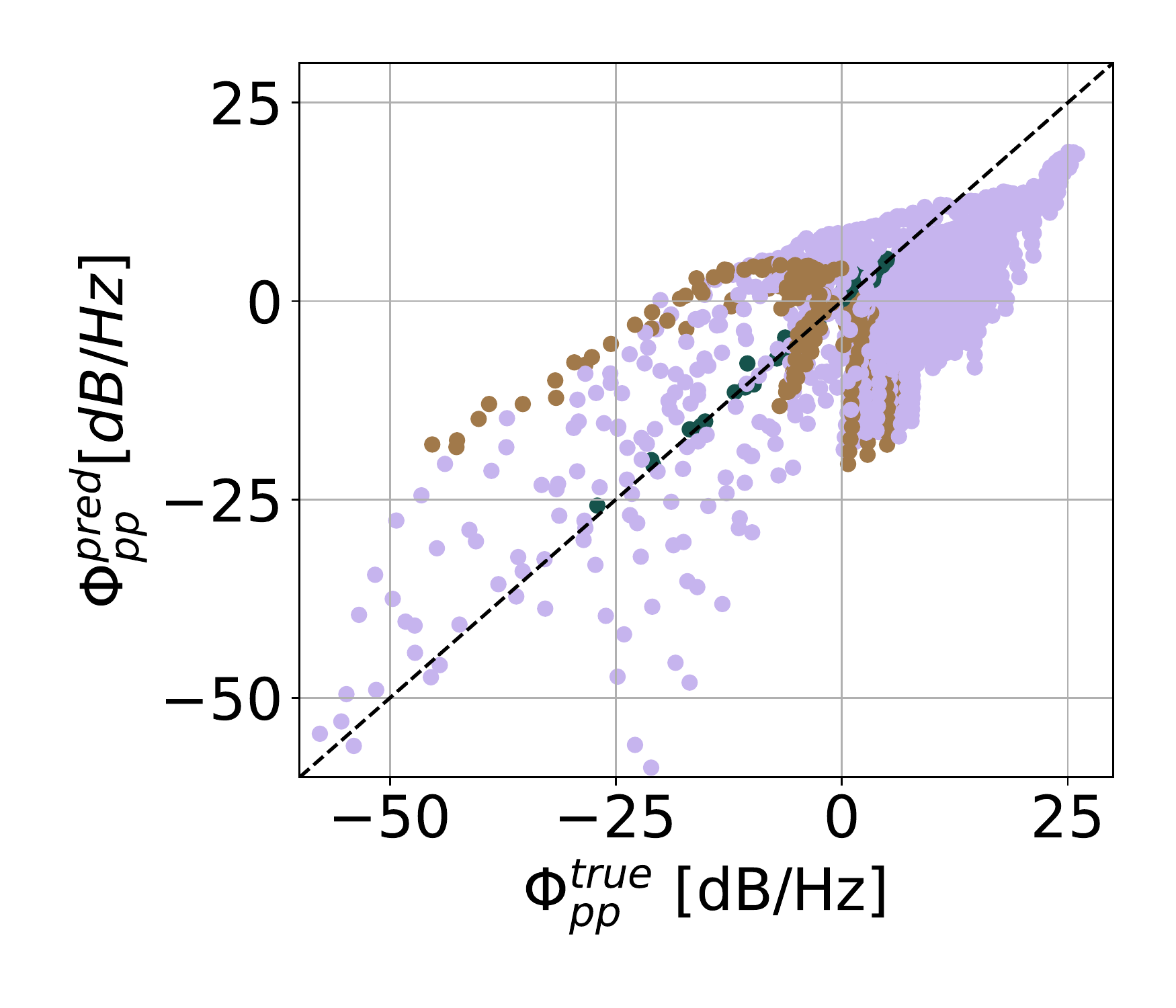}
    \caption{Lee lMSE = 74.89 dB/Hz}   
    \label{fig:perf_lee}
    \end{subfigure}
    \hfill
    \begin{subfigure}[b]{0.33\textwidth}   
    \centering 
    \includegraphics[width=\textwidth]{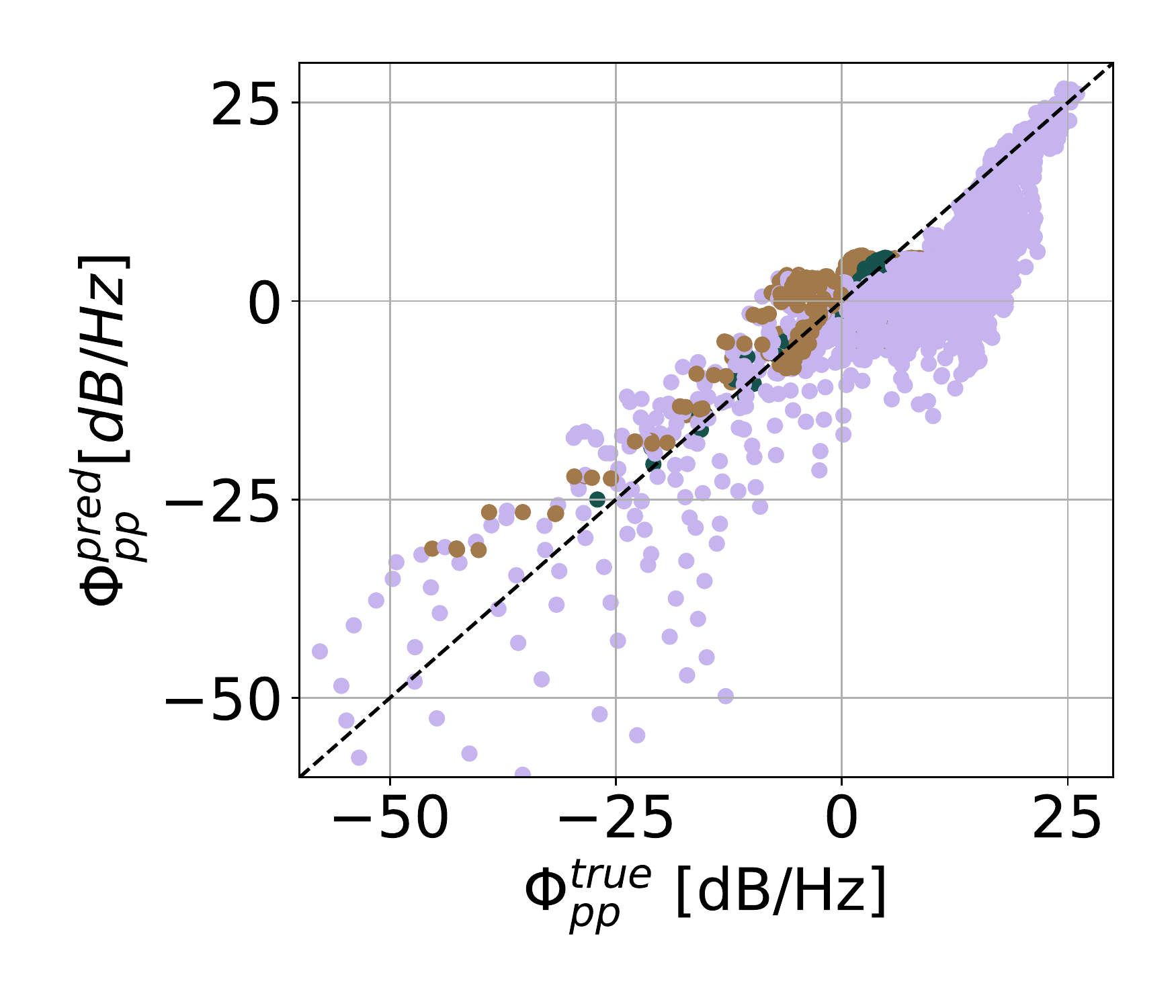}
    \caption{Dominique lMSE = 59.46 dB/Hz}  
    \label{fig:perf_dominique}
    \end{subfigure}
    \hfill
    \begin{subfigure}[b]{0.33\textwidth}   
    \centering 
    \includegraphics[width=\textwidth]{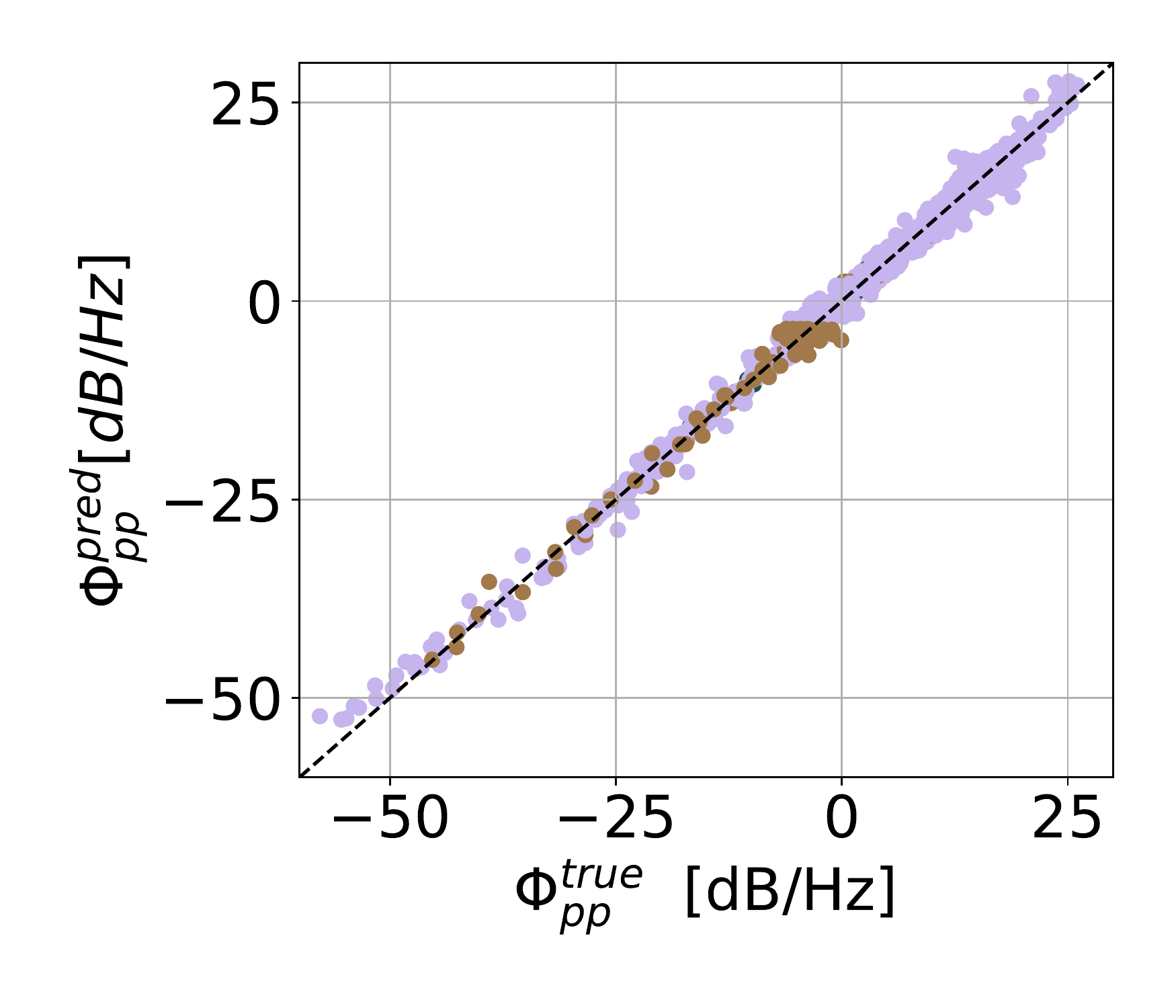}
    \caption{ANN lMSE = 0.88 dB/Hz}   
    \label{fig:perf_ANN}
    \end{subfigure}
    
    \caption{Global performance of the empirical models discussed in Sec \ref{sec:3}, on the dataset presented in Sec. \ref{sec:4}. The markers are \textcolor{color1}{\textbullet} for ZPG, \textcolor{color2}{\textbullet} for FPG and \textcolor{color3}{\textbullet} for APG, regardless of the dataset. For plotting purposes, only one point every ten is shown. } 
    
\label{fig:models_perf}
\end{figure*}

\begin{figure*}[p]
    \centering
    \begin{subfigure}[b]{0.45\textwidth}   
    \centering 
    \includegraphics[width=\textwidth]{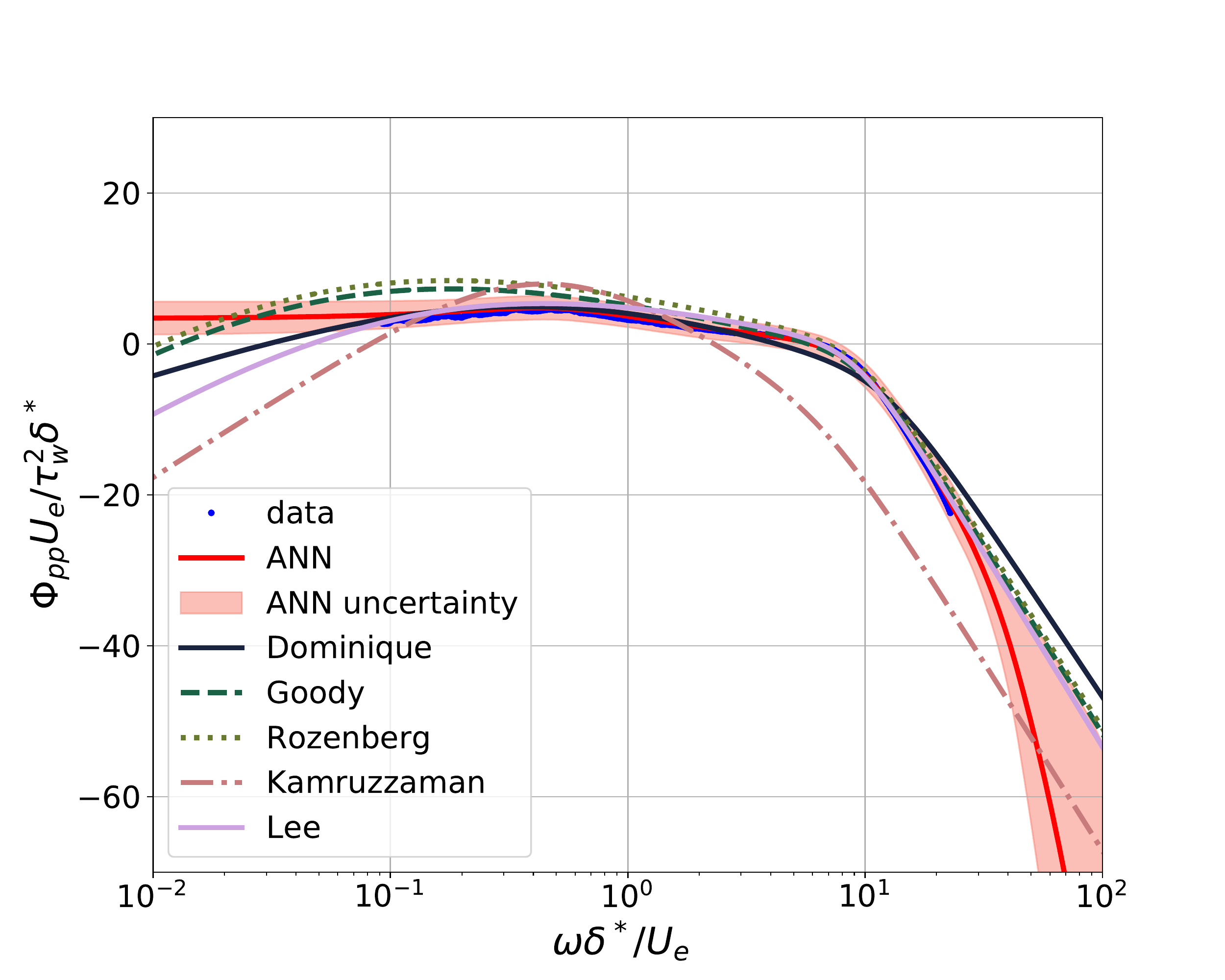}
    \caption{Salze's case with ZPG at $u_e = 45m/s$ ($\beta=0.0$, $R_T=13.11$, $C_f=2.64\times10^{-3}$, $\Pi=0.47$, $H=1.32$, $\Delta=7.8$, Ma = 0.133) }   
    \label{fig:NN Salze}
    \end{subfigure}
    \hfill
    \begin{subfigure}[b]{0.45\textwidth}  
    \centering 
    \includegraphics[width=\textwidth]{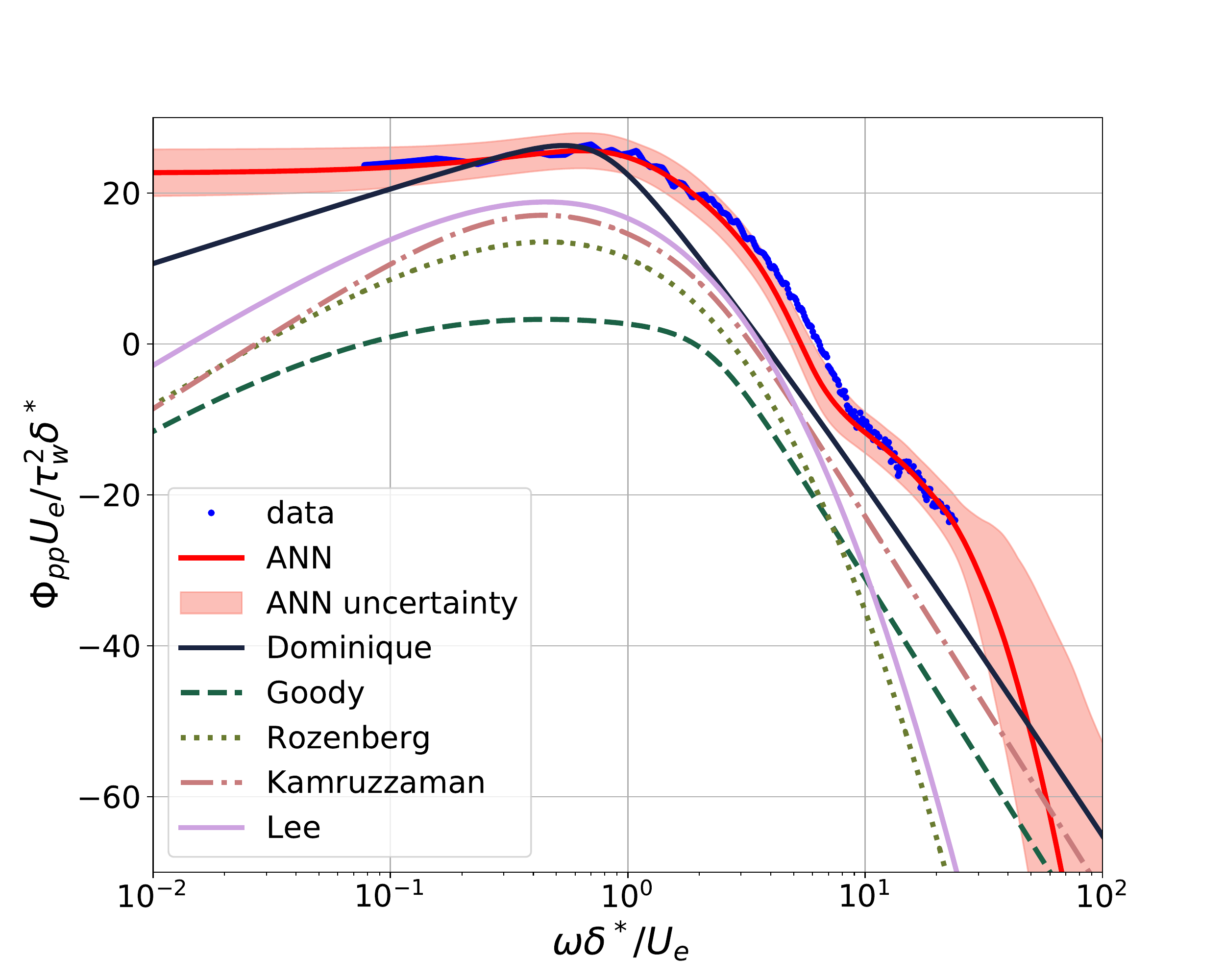}
    \caption{Deuse's case with APG at $x_c = 0.02$ ($\beta=15.59$, $R_T=2.58$, $C_f=1.02\times10^{-3}$, $\Pi=2.26$, $H=2.12$, $\Delta=3.1$, Ma= 0.211)}   
    \label{fig:NN_deuse}
    \end{subfigure}
    \vskip\baselineskip
\begin{subfigure}[b]{0.45\textwidth}
    \centering
    \includegraphics[width=\textwidth]{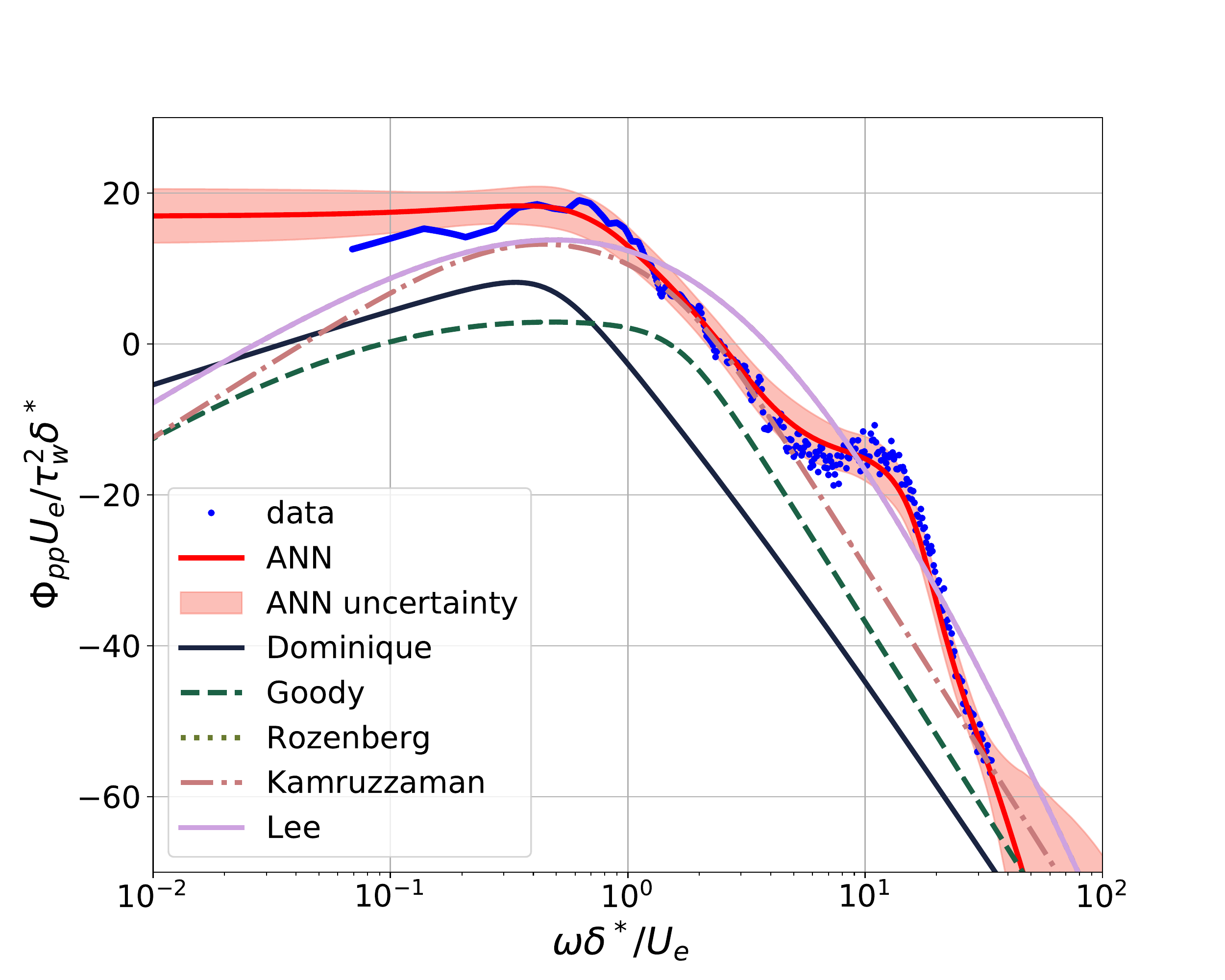}
    \caption{Hao's case with APG at $x_c = 0.02$ ($\beta=5.93$, $R_T=1.85$, $C_f=1.34\times10^{-3}$, $\Pi=2.26$, $H=2.15$, $\Delta=2.8$, Ma = 0.267)}
    \label{fig:NN_Hao}
    \end{subfigure}
    \hfill
        \begin{subfigure}[b]{0.45\textwidth}   
    \centering 
    \includegraphics[width=\textwidth]{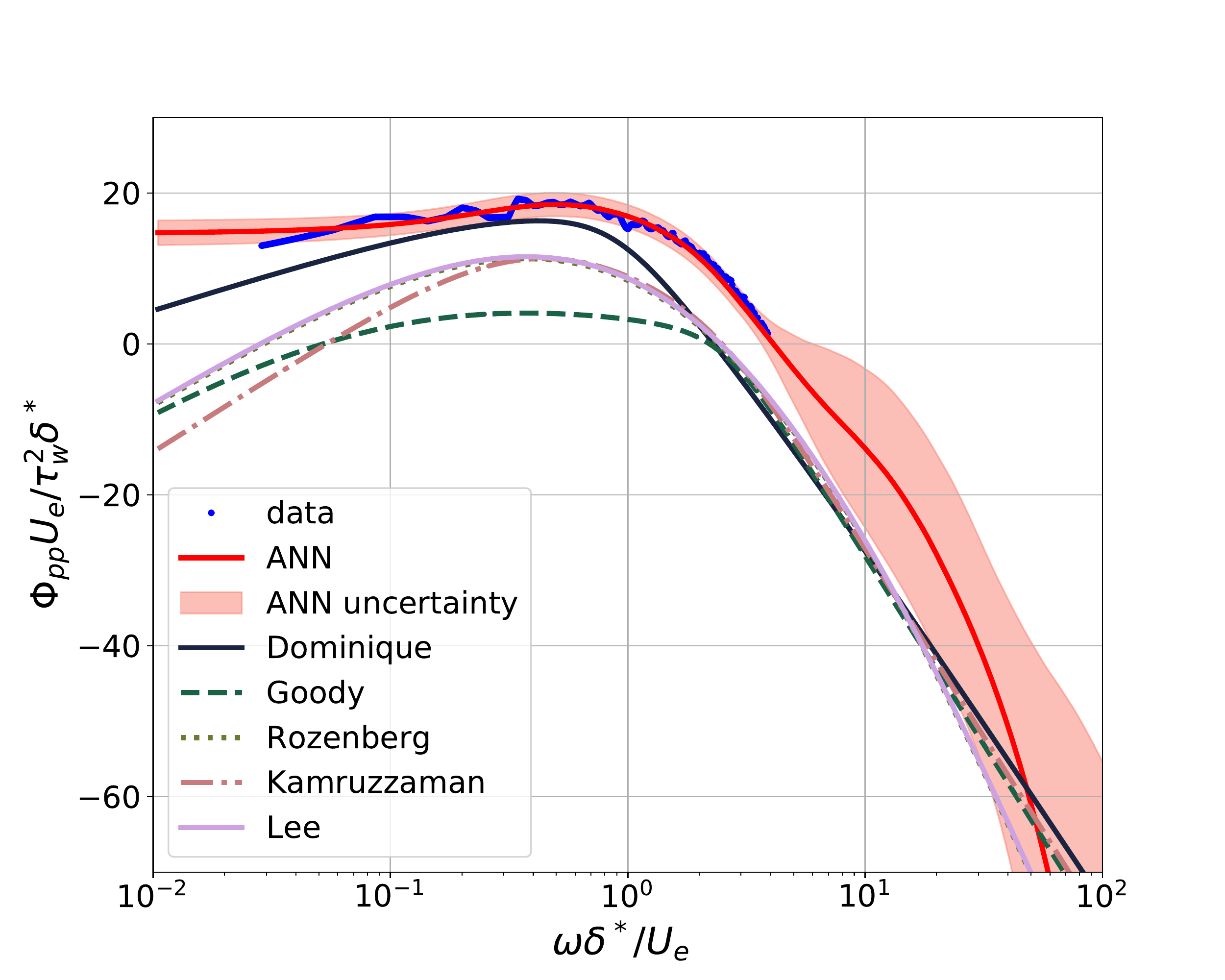}
    \caption{Christophe's case with APG at $x_c = 0.05$ ($\beta=6.32$, $R_T=X3.06$, $C_f=2.02\times10^{-3}$, $\Pi=1.28$, $H=1.82$, $\Delta=3.7$, Ma = 0.049)}   
    \label{fig:NN ch2}
    \end{subfigure}
    \vskip\baselineskip
    \begin{subfigure}[b]{0.45\textwidth}   
    \centering 
    \includegraphics[width=\textwidth]{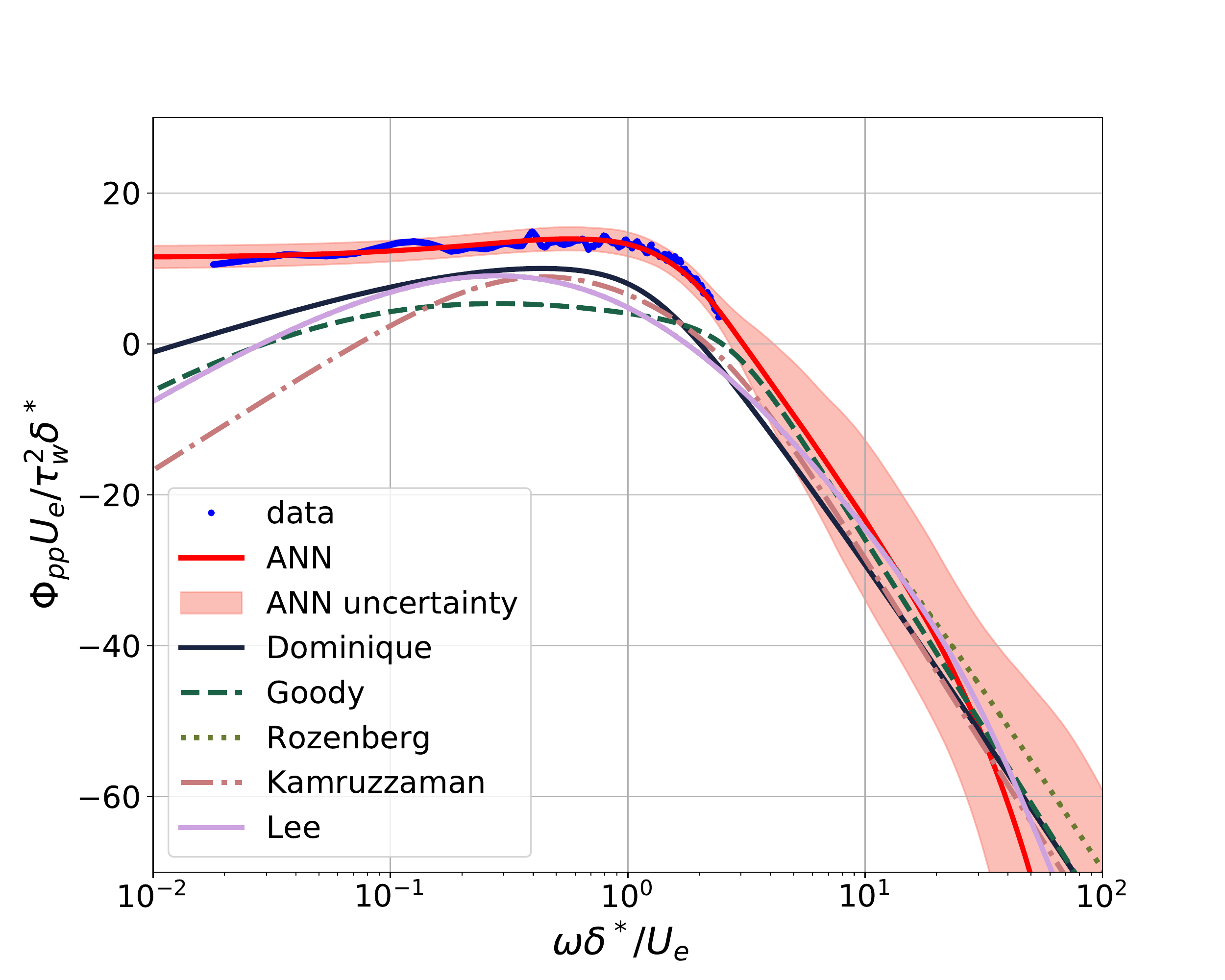}
    \caption{Christophe's case with APG at $x_c = 0.2$ ($\beta=3.79$, $R_T=3.47$, $C_f=3.13 \times 10^{-3}$, $\Pi=0.53$, $H=1.56$, $\Delta=5.0$, Ma = 0.053)}  
    \label{fig:NN ch1}
    \end{subfigure}
    \hfill
    \begin{subfigure}[b]{0.45\textwidth}   
    \centering 
    \includegraphics[width=\textwidth]{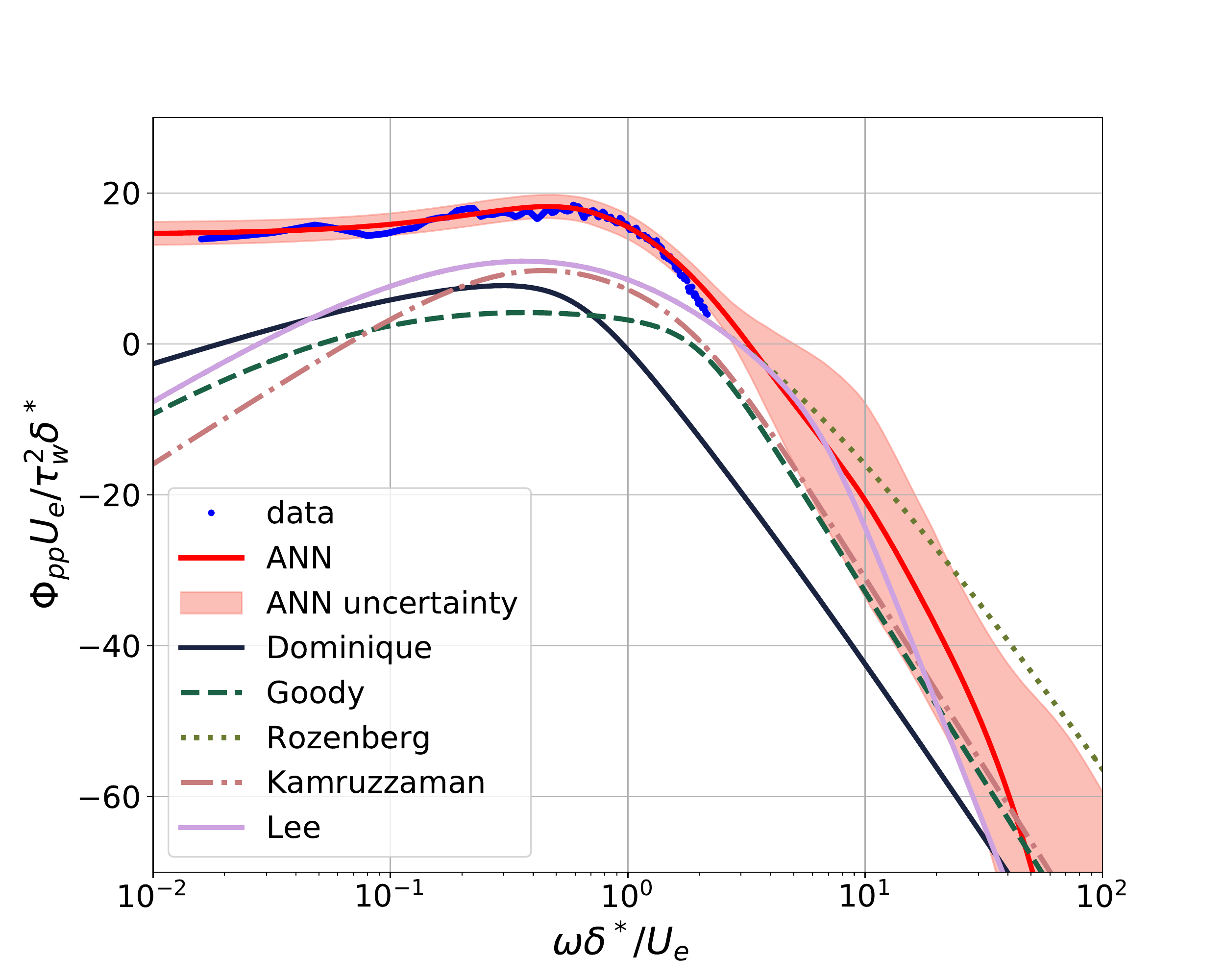}
    \caption{Christophe's case with APG at $x_c = 0.10$ ($\beta=3.37$, $R_T=2.33$, $C_f=2.52\times10^{-3}$, $\Pi=1.13$, $H=1.77$, $\Delta=3.8$, Ma= 0.051)}   
    \label{fig:NN ch3}
    \end{subfigure}
    \caption{Prediction of the ANN ensemble $\mu_{e}(\boldsymbol{x})$ form \eqref{eq:ANN_epistemic}, with confidence intervals $\mu\pm1.96\sigma(\boldsymbol{x})$, and comparison with selected semi-empirical models from Sec. \ref{sec:3} on the testing data.} 
\label{fig:NN_predictions}
\end{figure*}

\section{Results and Discussion}
\label{sec:6}

\subsection{Model Accuracy}\label{sec:6p1}

We provide a global performance score to all models discussed in Section \ref{sec:3}, and the new ANN model, on the available data. This score is defined as the average lMSE from \eqref{eq:MSE} over whole dataset. The global performance of all models is illustrated in Figure \ref{fig:models_perf}. A figure plots the prediction versus the available data for each model, and the caption reports the average lMSE error. The markers in the figure distinguish datasets in ZPG, APG and FPG.

Because of its inability to account for pressure gradients, Goody's model performs poorly in terms of global performance (average $lMSE=90.12$ dB/Hz) despite the excellent predictions for the ZPG conditions. More surprisingly, Kamruzzaman's model yields the largest errors (average $lMSE=160.98$ dB/Hz) despite being designed to account for pressure gradients, and having been shown to match well with Kamruzzaman's own experimental data \cite{kamruzzaman2015}. This discrepancy was also pointed out by Lee \emph{et al}\cite{lee2018} for both the CD airfoil and flat plate boundary layers. 

Rosenberg's (average $lMSE=76.14$ dB/Hz) and Lee's (average $lMSE=74.89$ dB/Hz) models yield similar global performances, gaining $20\%$ global lMSE over Goody's model. Much of the gain is observed in cases in APG. Their similar performance could also be partially attributed to the dataset used in this work: both of these models were calibrated to give similar predictions on the CD airfoil, but Lee's model was extended to account for various other cases that are not featured in the dataset collected in this work.

A further tangible improvement is obtained by Dominique's GEP model (average $lMSE=59.46$ dB/Hz), which performs better than the previous ones in both APG and FPG. Although this model has the same architecture as the previous models (a rational function in $\tilde{\omega}$), it accounts differently for pressure gradients (through $\beta$), introduces a function of $C_f$ and is independent of $\Pi$. The performance in APG conditions, however, remains limited.

Finally, the ANN's model shows a ten-fold gain in performance (average $lMSE=0.88$ dB/Hz) over the full range of conditions in the dataset. This shows the strength of the ANN as universal function approximator and offers some insights on the wall-pressure modeling problem: the success of the ANN model is reasonably explained by the fact that its functional relation is far more complex than the rational function assumed by all the other models. On the one hand, one might thus wonder whether the balance between model complexity and prediction performances should move the modeller's ingenuity beyond rational functionals. On the other hand, the comparatively poorer performances of the previous models, especially if compared to their performances on the data from which they were derived, shows that the dataset used in this work might not be sufficiently large and diverse to develop the most general ANN-based wall pressure spectra model. Nevertheless, the methodology described in this work could be easily extended, and the promising predictive capabilities of the ANN could be improved with more data.

We now move to the generalization capabilities of the ANN by testing it on the six profiles that were left out of the training. These were selected from all the available datasets and include one case in ZPG from Salze and five cases in APG, as these proved to be the most difficult to predict. The results are shown in Figure \ref{fig:NN_predictions}, with the caption recalling the dataset from which they are sampled and the input parameters. The prediction of all models is shown in the figure together with the ensemble mean prediction and confidence intervals for the ANNs. 
\begin{figure*}
\centering
\includegraphics[width=\textwidth]{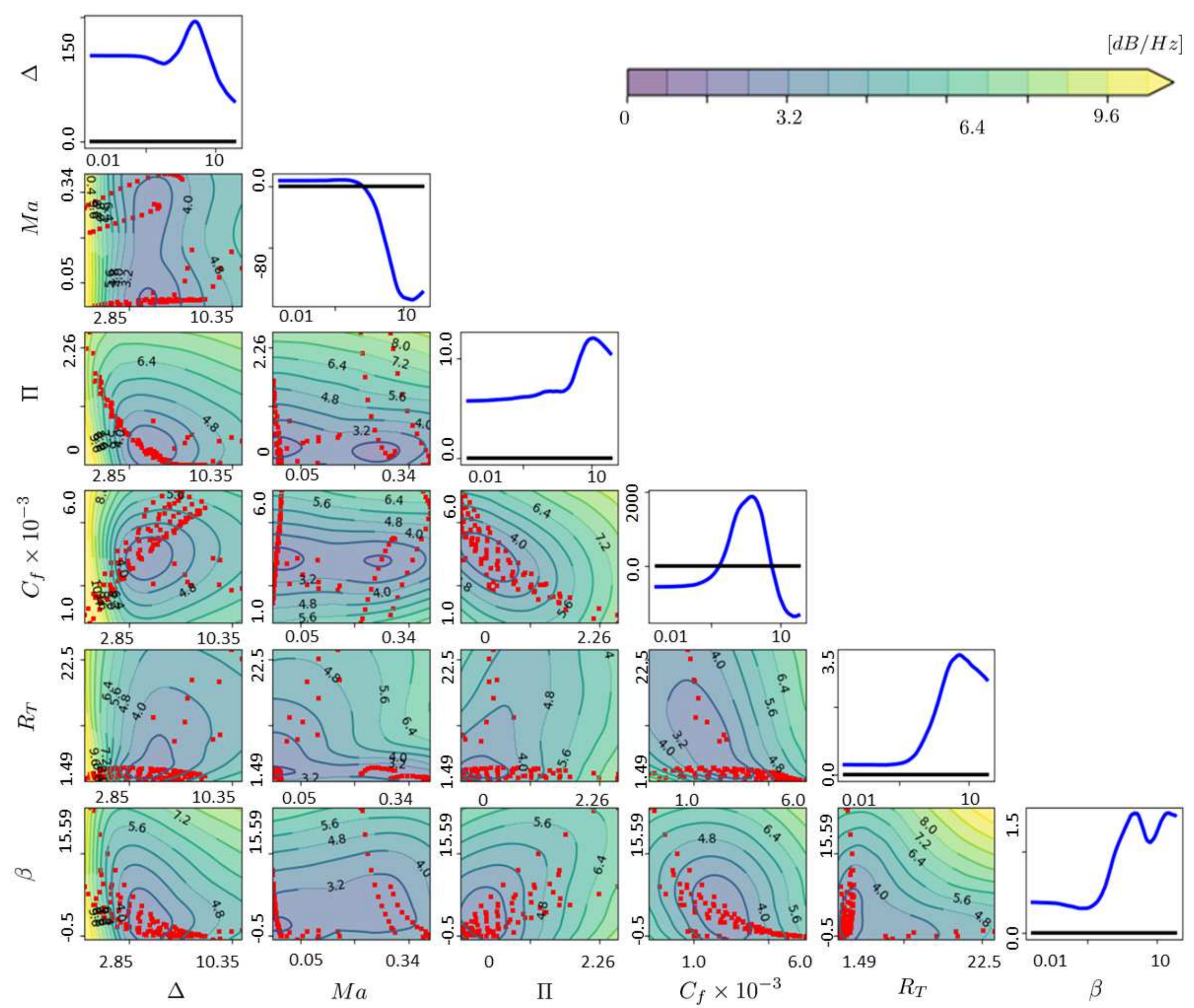}
\caption{Neural network uncertainties and average sensitivity via ensemble learning. For each combination of boundary layer parameters used as input, the contour map show the global measure of confidence $\mathcal{C}(\boldsymbol{x'})$ from \eqref{eq:ANN_confidence}, varying only two parameters and keeping the others at their average value. The red dots are the training points from the database. The diagonal shows the average sensitivity $\mathcal{S}_{\boldsymbol{x}'_i}$ from  \eqref{S_omega} with respect to the input parameter $\boldsymbol{x}_i$, as a function of the dimensionless frequency $\tilde{\omega}$. The line with  $\mathcal{S}_{\boldsymbol{x}'_i}=0$ is highlighted in black for plotting purposes. }
\label{fig:NN_accuracy}
\end{figure*}

For the first profile in \ref{fig:NN Salze}, in ZPG, all models yield good predictions except for Kamruzzaman's model. This model misses the roll-off high-frequency point, presumably because of its different dependency on the Reynolds number. The ANN provides an excellent fit within its training frequency range, but departs from the data and from Goody's model at both high and low frequencies. As expected, Rozenberg's closely match Goody's model in this condition, but both models slightly overpredict Salze's data at low frequencies. The models of Lee and Dominique also match closely Goody's model prediction while providing a better fit at low frequencies.

Large uncertainty band are observed for the ANN for some conditions, e.g. for $\omega \delta^*/u_e >25$ in ZPG. They can be explained by the lack of training data. In these cases, the ANN predictions are biased by the few profiles that include data in this frequency range. These are the profiles from Hao (see also Fig. \ref{fig:NN_Hao}), which are in APG conditions and feature a steeper decay.

The extrapolation problem is less severe for Dominique's GEP model because its rational function form keeps the expected trends outside the training range. The trade-off between model complexity and model performance is a classic problem in the machine learning literature\cite{Abu} and its resolution for wall pressure spectra modelling requires a larger dataset. This suggests a possible hybrid approach where  the rational function remains imposed, and the ANN is used to predict the necessary coefficients.

The complexity of the ANN model pays off in the APG conditions, as clearly shown by the test case from Deuse's (Fig. \ref{fig:NN_deuse}) and Hao's datasets (Fig. \ref{fig:NN_Hao}). These are the test cases for which data is available at the highest frequencies and exhibit wall pressure spectra that rational functions don't seem to describe well. This is particularly true for the test case in Hao's datasets (Fig. \ref{fig:NN_Hao}), characterized by a hump at high frequencies. Although the prediction of Lee's model shows a remarkable agreement with the data, the ANN is the only model capable of predicting such a trend. The same conclusions hold for the test case in Deuse's dataset (Fig. \ref{fig:NN_deuse}), which also feature a gentle change in the slope of the power spectra at $\tilde{\omega}\approx 10$. In this case, Lee's model is less accurate, and the ANN approach largely outperforms all empirical models.

Due to the lack of high-frequency data in Christophe's strong APG database, the ANN is asked to extrapolate over a significant frequency range to predict those spectra. Relatively large uncertainty bands are indeed observed above $\tilde{\omega} \simeq 4$. Such extrapolation seems to be in reasonable agreement with the empirical models for the test case in \ref{fig:NN ch1} but not for the ones in Figs \ref{fig:NN ch2} and \ref{fig:NN ch3}. On the other hand, the accuracy at the lowest frequency is remarkable and unmatched by the any of the semi-empirical models.

Finally, most of the points in the test set fall within the ANN confidence intervals. This validates the ensemble methodology for uncertainty quantification.

\subsection{Uncertainties and Sensitivities}\label{sec:6p2}

We analyze the confidence intervals and the sensitivity of the ANN ensemble over the full range of available data. The goal was to identify regions where additional data would be mostly beneficial to improve the model and to analyze the sensitivity of the confidence bounds to the input parameters. 

Let $\sigma_E(\boldsymbol{x'},\tilde{\omega})$ be the standard deviation of the ANN ensemble from \eqref{eq:ANN_epistemic}, as a function of the dimensionless frequency $\tilde{\omega}$, for a given input in $\boldsymbol{x'}=[\Delta, Ma, \Pi, C_f, R_T, \beta ]$. Let $[\omega_1,\omega_2]$ be the largest range of available dimensionless frequencies, which in this work is $[0.01,0.35]$. For every input $\boldsymbol{x'}$, we define a global measure of confidence as 
\begin{equation}\label{eq:ANN_confidence}
    \mathcal{C}(\boldsymbol{x'})=\frac{1}{\tilde{\omega}_2-\tilde{\omega}_1}\int^{\tilde{\omega}_2}_{\tilde{\omega}_1} \sigma_E(\boldsymbol{x'},\tilde{\omega}) d\tilde{\omega}\,.
\end{equation}

The contour-plot of the confidence measure $\mathcal{C}$ is shown in Figure \ref{fig:NN_accuracy} in the form of a scatter matrix on the planes defined by pairs of inputs (e.g. $\Delta$ versus $Ma$, $\delta$ versus $\Pi$, etc.). The contour plot shows the value of $\mathcal{C}$ together with the available data, shown as a scatter plot.

The diagonal of the scatter matrix shows the averaged sensitivity to each input parameter as a function of the dimensionless frequency $\tilde{\omega}$. In other words, given $y=10\log_{10}(\tilde{\Phi}_{pp}(\tilde{\omega},\boldsymbol{x}))$ the output of the ANNs for an input $\boldsymbol{x}'$ and a frequency $\tilde{\omega}$, the global (averaged) ensemble sensitivity is defined as 

\begin{equation}
    \label{S_omega}
    \mathcal{S}_{\boldsymbol{x}'_i} (\tilde{\omega})=\mathbb{E}_{\sim E, \boldsymbol{X}} \{ \partial_{\boldsymbol{x}'_i} y\}\,,
\end{equation} where the expectation is computed over both the ensemble ($E$) and the full dataset ($\boldsymbol{X}$) and the partial derivative $\partial_{\boldsymbol{x}'_i} y$ with respect to the input $\boldsymbol{x}'_i$ is computed analytically via back-propagation.

The contour plots confirm high uncertainty in regions lacking data. The region of highest uncertainties are observed at large $R_T$, large $\beta$ and low $\Delta$. Of these three parameters, the uncertainties grow more significantly with decreasing $\Delta$. The figures in the diagonal show that this last parameter has a strong impact at low frequency, while others, such as $Ma$ and $R_T$, mostly play at high frequencies. It is worth noticing that the averaged value of the derivative with respect to $R_T$ is $\partial_{R_T} y  \approx 3.5$ at high frequencies, in agreement with the expected dependency of the Reynolds number in Goody's model which should averaged at 4.0. Moreover, the averaged sensitivity $\mathcal{S}_{\boldsymbol{x}'_i} (\tilde{\omega})$ is always positive for ${\boldsymbol{x}'_i}=\Pi,\,R_T, \,\beta$ as it is the case in other empirical models, while the sensitivity changes sign for ${\boldsymbol{x}'_i}=C_f, Ma$.

\begin{figure}[h]
\includegraphics[width=0.45\textwidth]{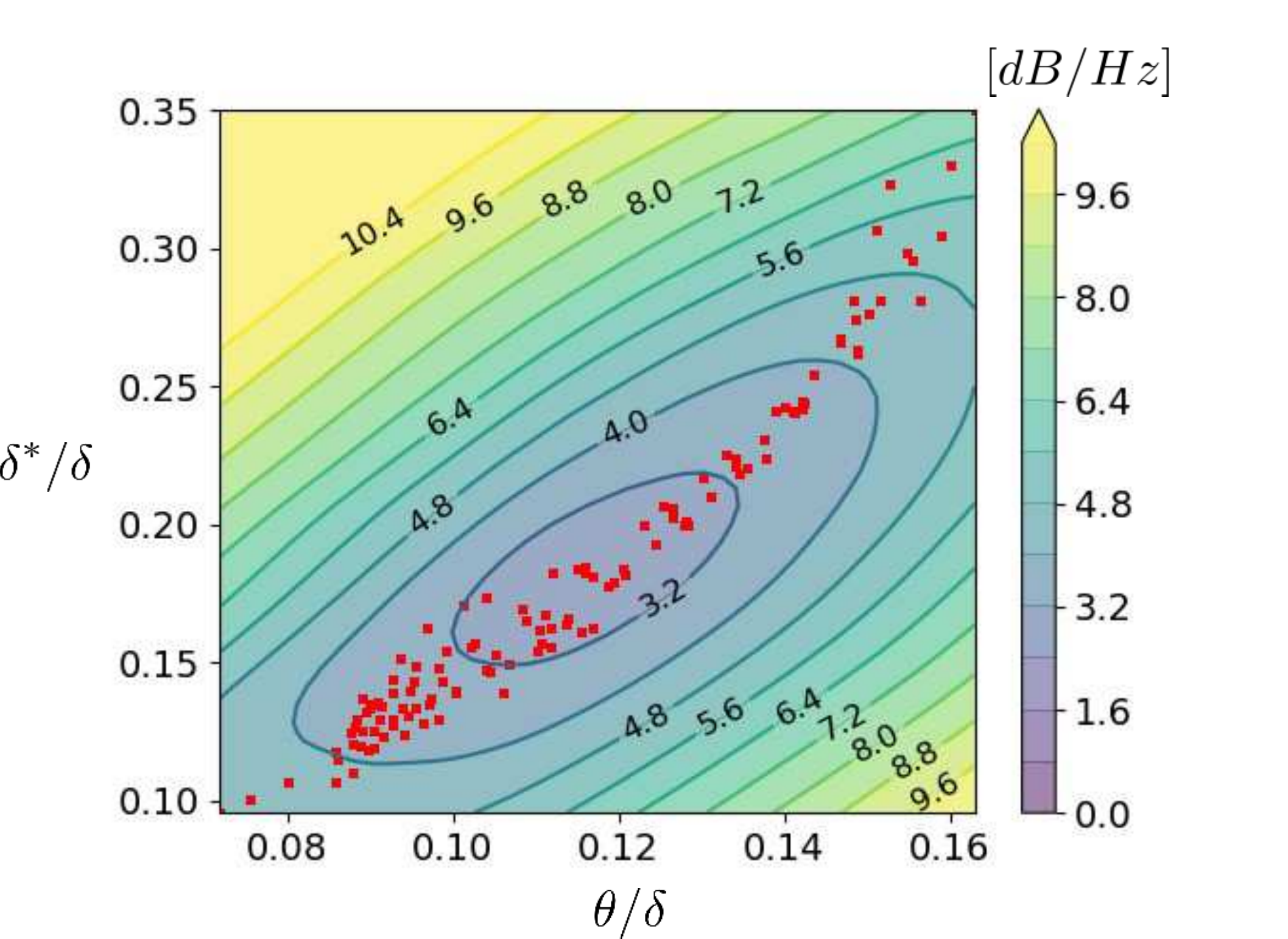}
\caption{Correlation between the variables $\delta$, $\delta^*$ and $\theta$. The red dots are the training points from the database.  The contour plots display display the global measure of confidence $\mathcal{C}(\boldsymbol{x'})$ from \eqref{eq:ANN_confidence} for $\delta^*/\delta$ and $\theta/\delta$, while the others parameters are fixed to their mean averaged value.}
\label{fig:NN_correlation}
\end{figure}

Some of the regions of high uncertainties (e.g. large $R_T\sim 20$ \emph{and} $\beta\sim 10$) are unexplored territory for which data would be particularly valuable in improving the model performances. Others are less interesting because physically inaccessible due to the correlation between inputs. For example, the plane $\Pi-\Delta$ reveals a strong correlation between these two variables, and an apparent impossiblity for the boundary layer to exhibit high values for both simultaneously. The corresponding region of high uncertainty is thus not a concern. Strongly correlated variables were used in this work as a way to increase the predictive capabilities of the model. This is the case for example of $\Delta$ and $H$, which contains the parameters $\delta$, $\delta^*$ and $\theta$. These are almost linearly correlated in the training data used in this work, as shown by the scatter plot in Figure \ref{fig:NN_correlation} together with the contour of $\mathcal{C}$. Therefore, the regions of high uncertainties in this plot do not seem physically accessible. At least within the range of parameters investigated, one might consider removing one of those two parameters in future modeling efforts.

\section{Conclusion}
\label{sec:7}

Existing semi-empirical models fail to predict wall pressure spectra for strong adverse pressure gradients, and data-driven models such as neural networks offer an attractive alternative. The artificial neural network is trained on a dataset constituted of three high-fidelity numerical simulations on the CD-airfoil and one set of experimental measurements on a flat plate boundary layer. The methodology used to extract the relevant boundary layer parameters required by the model was extended to account for large pressure gradients encountered in industrial applications and the CD-airfoil simulations. The neural network constructed in this work provided a significant improvement over the other existing models in those conditions. Finally, we investigated the uncertainties of the neural network predictions and the sensitivity to its inputs using an ensemble approach. Those allow for identifying regions where new training data would be most beneficial to the model's accuracy.


In contrast with semi-empirical models assuming a functional dependence (in this instance, a ratio of polynomials), which can filter out outliers to some extent, data-driven methods rely entirely on the quality of their inputs. Therefore, an elaborate methodology has been necessary for computing the parameters of boundary layers subjected to an adverse pressure gradient. This method is based on the theoretical developments of Nickles \cite{nickels} and Nagib \cite{nagib}, about the non-universality of the von Karman coefficient. This procedure showed that the pressure gradient strongly impacts the von Karman coefficients of the log law. We accounted for these changes while computing the boundary layer parameters for three high-fidelity numerical simulations on the CD-airfoil and one set of experimental measurements on a flat plate boundary layer.

Methods for data-driven regression such as genetic programming and neural network allow for deriving complex  nonlinear functions from data but require an appropriate set of input parameters. A dimensional analysis using the Buckingham-Pi theorem yielded a set of eight dimensionless numbers to model the wall pressure spectra. We verified that this set of parameters was sufficient to model the present dataset using a manifold mapping approach for dimensionality reduction. We observed that the available dataset doesn’t contain samples with identical boundary layer parameters leading to different wall pressure spectra. While exploring the dataset, we observed a clear correlation between the  Zagarola-Smits’s parameter $\Delta$ and the shape factor $H$. If this correlation proved true on extended data, this suggests that both those parameters are unnecessary, and we could consider removing one.


The database investigated in this work provides a thorough investigation for the CD airfoil under a broad range of inflow parameters, including many strong adverse pressure gradient boundary layers. Among all the empirical models presented in this work, Lee's and Rozenberg's model provided the best match to the data. However, even those model tends to underpredict the exact amplitude of the spectra by 5 to 10 dB/Hz. 
Dominique's data-driven model using GEP slightly improves over Lee's predictions in some cases, but this improvement is not consistent for all turbulent boundary layers. In addition, it does not capture the correct roll-off frequency for the transition toward the high-frequency spectral decay. The imperfections of the GEP model are the result of the convergence difficulties of Genetic programming approaches when the parameter space formed by the number of operators and input functions becomes too large. 

The neural network constructed in this work provided a significant improvement over the other models, with accurate predictions globally within $\pm 1 dB/Hz $ accuracy for the different flow cases. However, the model tends to feature a steeper decay at high frequencies due to the limited training data. 

An analysis of the uncertainties through ensemble learning highlighted the lack of data for large pressure gradients and medium to large Reynolds numbers or low Zagarola-Smits’s parameter. This caused an increase in the model uncertainty for such flow configurations and indicated this area of parameters as a promising region for additional data in the future. In addition, different airfoil geometries should be added to the dataset as it is currently mainly composed of simulations over the CD airfoil.

The authors conjecture that this approach could be extended to a broader range of boundary layer parameters without additional modifications in the long term. This would make a great tool to propose models that perform well on a larger dataset, including different physical assumptions (e.g. pressure gradient, compressibility effects, etc.)  without requiring additional effort on the methodology front.

\begin{acknowledgments}

The French global automotive Valeo partly supported this research in the context of a thesis for the investigation of noise induced by HVAC systems. The authors thank Dr. Salze and Pr. Bailly from Ecole Centrale de Lyon, Ing. Deuse and Pr. Sandberg from the University of Melbourne, Dr. Hao and Pr. Moreau from Université de Sherbrooke and Dr. Christophe from von Karman Institute for providing the experimental and numerical data.
\end{acknowledgments}

\appendix

\section{ List of symbols }\label{app:symbol}

\begin{table}[h]
\begin{ruledtabular}
\begin{tabular}{llr}
Name & Symbol & Definition\\
\hline
wall pressure spectra & $\Phi_{pp}$ & Eq. \ref{eq:Phipp} \\
wavenumber & $\omega$ & $2\pi f$ \\
equilibrium velocity & $u_e$ & Maximum (pseudo) velocity \\
boundary layer thickness & $\delta$ & $y\left(u=0.99u_e\right)$\\
displacement thickness & $\delta^*$ & Eq. \ref{eq:thickness} a\\
momentum thickness & $\theta$ & Eq. \ref{eq:thickness} b\\
viscous length scale & $\delta_\nu$ & $\nu / u_\tau$ \\
density & $\rho$ & Ideal gas law \\
viscosity & $\nu$ & Sutherland’s law \\
wall shear stress & $\tau_w$ & $\rho u_\tau^2$\\
friction velocity & $u_\tau$ & Fig. \ref{fig:bl_model}\\
speed of sound & $c_0$ & 343 m/s \\
wake parameter & $\Pi$ & Eq. \ref{eq:wake} \\
normalize frequency & $\tilde{\omega}$ & $\omega \delta^* / u_e$ \\
shape factor & $H$ & $\delta^*/\theta$ \\
Zagarola-Smits's parameter & $\Delta$ & $\delta/\delta^*$ \\
Mach number & $M$ & $u_e/c_0$ \\
friction coefficient & $C_f$ & $\tau_w / \rho u_e^2$ \\
Reynolds number & $R_T$ & $(\delta^*/u_e)/(\nu/u_{\tau}^2)$ \\
Clauser parameter & $\beta$ & $(\theta/\tau_w)\partial p/\partial x$ \\
wall unit velocity & $u^+$ & $u/u_\tau$ \\
wall unit distance & $y^+$ & $y u_\tau / \nu$ \\
von Karman slope & $\kappa$ & Eq. \ref{eq:nickel_kappa} \\
von Karman constant & $B$ & Eq. \ref{eq:nagib} \\
inner pressure gradient & $p_x^+$ & $(\nu/(\rho u_\tau^3) \partial p_x$ \\
critical distance & $ y_c$ & Eq. \ref{eq:nickel_yc} \\
wall unit critical distance & $y_c^+$ & $y_c u_\tau / \nu$ \\
\end{tabular}
\end{ruledtabular}
\end{table}

\section{Database range}

\begin{table*}
\caption{\label{tab:input}Boundary layer parameters for the four analyzed databases.}
\begin{adjustbox}{max width=\textwidth}
\begin{tabular}{c|c|c|c|c|c|c|c|c|c|c}
database & N & W & $\tilde{\omega} = \omega \delta^*/u_e$ & $\Delta = \delta/\delta^* $ & $H = \delta^*/\theta$ & $M=u_e/c_0$ & $\Pi$ & $C_f=\tau_w/\rho u_e^2$ & $R_T = (\delta^*/u_e)/(\nu/u_{\tau}^2)$ & $\beta = (\theta/\tau_w)\partial_x p $ \\
 \hline
 Salze (2014) & 10 & 1 & $0.03 \rightarrow 33.6$ & $6.14 \rightarrow 10.48$ & $0.07 \rightarrow 0.12$ & $0.07 \rightarrow 0.17$ & $0.03 \rightarrow 0.75$ & $0.002 \rightarrow 0.004$ & $7.60 \rightarrow 22.5$ & $-0.47 \rightarrow 0.57$ \\
 Deuse (2020) & 13 & 0.77  & $0.01 \rightarrow 24.2$ & $3.10 \rightarrow 6.48$ & $0.11 \rightarrow 0.15$ & $0.21\rightarrow 0.27$ & $0.14 \rightarrow 2.26$ & $0.001 \rightarrow 0.004$ & $2.58 \rightarrow 3.54$ & $0.36\rightarrow 15.59$  \\
 Hao (2020) & 16 & 0.62 & $0.02 \rightarrow 34.4$ & $2.85 \rightarrow 7.57$ & $0.09 \rightarrow 0.16$ & $0.27 \rightarrow 0.34$ & $0.0 \rightarrow 2.2$ & $0.001 \rightarrow 0.006$ & $1.56 \rightarrow 2.25$ & $-0.15 \rightarrow 5.93$  \\
 Christophe (2011) & 78 & 0.13 & $0.002 \rightarrow 4.65$ & $3.28 \rightarrow 8.72$ & $0.86 \rightarrow 0.16$ & $0.05 \rightarrow 0.06$ & $0.0 \rightarrow 1.43$ & $0.002 \rightarrow 0.006$ & $1.49 \rightarrow 3.48$ & $-0.05 \rightarrow 7.94$  \\
 \hline
 \hline
 All & \multicolumn{2}{c|}{117} & $0.01 \rightarrow 34.4$ & $2.85 \rightarrow 10.48$ & $1.23 \rightarrow 2.15$ & $0.05 \rightarrow 0.34$ & $0 \rightarrow 2.26$ & $0.001 \rightarrow 0.006$ & $1.49 \rightarrow 22.5$ & $-0.47 \rightarrow 15.59$  \\
\end{tabular}
\end{adjustbox}
\end{table*}

\bibliography{aipsamp}

\end{document}